\documentclass[journal]{IEEEtran}

\newif\ifreview

\reviewfalse
\usepackage{tabularx}
\usepackage[table]{xcolor}
\usepackage{array}
\usepackage{amsmath,amsfonts}
\usepackage{algorithmic}
\usepackage{algorithm}
\usepackage{array}
\usepackage[caption=false,font=footnotesize,listofformat=subsimple,labelformat=simple,subrefformat=simple]{subfig}

\makeatletter
\renewcommand{\p@subfigure}{\thefigure}
\makeatother
\usepackage{textcomp}
\usepackage{stfloats}
\usepackage{url}
\usepackage{verbatim}
\usepackage{graphicx}
\usepackage{cite}
\usepackage{makecell}
\begin{document}
%
% paper title
% Titles are generally capitalized except for words such as a, an, and, as,
% at, but, by, for, in, nor, of, on, or, the, to and up, which are usually
% not capitalized unless they are the first or last word of the title.
% Linebreaks \\ can be used to get better formatting as desired.
% Do not put math or special symbols in the title.
\title{Ultra-Wideband Stacked Patch Antenna for \\ D-Band Applications}

%\title{Concurrent Dual-mode Coupler for a Multi-mode Multi-drop Link Over Substrate-integrated Waveguides
%}
%Concurrent Multi-Mode Excitation for Mode Division Multiplexing over
%Substrate Integrated Waveguides

%
%
% author names and IEEE memberships
% note positions of commas and nonbreaking spaces ( ~ ) LaTeX will not break
% a structure at a ~, so this keeps an author's name from being damaged across
% two lines.
% use \thanks{} to gain access to the first footnote area
% a separate \thanks must be used for each paragraph as LaTeX2e's \thanks
% was not built to handle multiple paragraphs
%

%
% For Reviewing
%
\newcommand{\remove}[1]{\ifreview \textcolor{red}{\sout{#1}} \fi}
\newcommand{\lastrev}[1]{\ifreview \textcolor{blue}{#1}\else{#1}\fi}

\author{Mohamed~Elsawaf,~\IEEEmembership{Graduate Student Member,~IEEE}, Neelam Prabhu Gaunkar, ~\IEEEmembership{Member,~IEEE}, \\Georgios C. Dogiamis,~\IEEEmembership{Member,~IEEE}
        and~Constantine~Sideris,~\IEEEmembership{Senior Member,~IEEE}% <-this % stops a space
\thanks{Manuscript received  \emph{(Corresponding author: Constantine Sideris)}. This work is funded by Intel Corporation. Mohamed Elsawaf is with the Department of Electrical and Computer Engineering, University of Southern California, Los Angeles, CA, 90089 USA. Neelam Prabhu Gaunkar is with Intel Corporation, Technology Research, Chandler, Arizona. Georgios C. Dogiamis was with Intel Corporation and is now with Deca Technologies, Tempe, Arizona. Constantine Sideris is with the Department of Electrical Engineering, Stanford University, Stanford, CA 94305 (e-mail: sideris@stanford.edu). }% <-this % stops a space
}

% note the % following the last \IEEEmembership and also \thanks - 
% these prevent an unwanted space from occurring between the last author name
% and the end of the author line. i.e., if you had this:
% 
% \author{....lastname \thanks{...} \thanks{...} }
%                     ^------------^------------^----Do not want these spaces!
%
% a space would be appended to the last name and could cause every name on that
% line to be shifted left slightly. This is one of those "LaTeX things." For
% instance, "\textbf{A} \textbf{B}" will typeset as "A B" not "AB." To get
% "AB" then you have to do: "\textbf{A}\textbf{B}"
% \thanks is no different in this regard, so shield the last } of each \thanks
% that ends a line with a %, and do not let a space in before the next \thanks.
% Spaces after \IEEEmembership other than the last one are OK (and needed) as
% you are supposed to have spaces between the names. 
% This is a minor point as most people would not even notice if the said evil
% space somehow managed to creep in.

% The paper headers
\markboth{IEEE Transactions on Antennas and Propagation}%
{Shell \MakeLowercase{\textit{et al.}}: Bare Demo of IEEEtran.cls for IEEE Journals}
% The only time the second header will appear is for the odd-numbered pages
% after the title page when using the two-side option.
% 
% *** Note that you probably will NOT want to include the author's ***
% *** name in the headers of peer review papers.                   ***
% You can use \ifCLASSOPTIONpeerreview for conditional compilation here if
% you desire.

% If you want to put a publisher's ID mark on the page, you can do it like
% this:
%\IEEEpubid{0000--0000/00\$00.00~\copyright~2015 IEEE}
% Remember, if you use this, you must call \IEEEpubidadjcol in the second
% column for its text to clear the IEEEpubid mark.

% used for special paper notices
%\IEEEspecialpapernotice{(Invited Paper)}

% Make the title area
\maketitle

% As a general rule, do not put math, special symbols, or citations
% in the abstract or keywords.
\begin{abstract}
This article proposes a wideband stacked dual-patch antenna designed for D-band applications. The antenna is demonstrated with both wireless and dielectric waveguide (DWG) channels. The patch dimensions and spacings are chosen to ensure that the $10~dB$ return loss (RL) bandwidth covers the entire D-band ($110$--$170~GHz$). The antenna is fabricated on a seven-layer organic packaging technology. The patches are fabricated on layers 1 and 3 within a shielding cavity formed by closely spaced vias of $< \lambda/20$ pitch. The fabrication technology supports direct chip-level excitation of the antenna using a flip-chip packaging interface, and the design methodology for this feeding approach is outlined in the paper. In the fabricated design, however, a transition to the WR6 metallic waveguide with ($<~1~dB$) insertion loss (IL) is incorporated for feeding. This facilitates alignment and measurement of the structure using external test equipment. The full antenna and transition are integrated in a back-to-back configuration, achieving a measured $10~dB$ return loss bandwidth of more than $60~GHz$ ($110$--$170~GHz$) for the wireless link, thereby covering the entire D-band. To the best of our knowledge, this is the widest antenna bandwidth reported in the same frequency range. The measured peak gain is $8.5~dBi$ ($35^\circ$ H-plane $3~dB$ beamwidth) with a gain variation of $3.5~dBi$ within the operating bandwidth. For the dielectric waveguide configuration, the measured $10~dB$ RL bandwidth also exceeds $60~GHz$, fully encompassing the D-band. The simulated insertion loss coupling into the dielectric waveguide is $3.5~dB$. A 1-meter dielectric waveguide channel fed with the proposed antenna is measured. The overall channel exhibits a total insertion loss of $14.7~dB$, of which a simulated $3.5~dB$ at $140~GHz$ is attributed to the loss of each side that feeds the DWG.
\end{abstract}

% Note that keywords are not normally used for peer-reviewed papers.

% For peer review papers, you can put extra information on the cover
% page as needed:
% \ifCLASSOPTIONpeerreview
% \begin{center} \bfseries EDICS Category: 3-BBND \end{center}
% \fi
%
% this IEEEtran command inserts a page break and
% creates the second title for peer review papers
% creates the second title. It will be ignored for other modes.
\IEEEpeerreviewmaketitle

\section{Introduction} 
\IEEEPARstart{A}{s} the demand for higher data-rates grows, systems are required to operate at higher and higher frequencies. In recent years, sub-THz technologies have attracted a lot of attention~\cite{IEEE_network,survey,Reynaert,Disperison_limitation} for their potential application in the next generation of wireless and wireline communication technologies. One of the key bands for sub-THz communications is the D-band ($110$--$170~GHz$), where conventional CMOS technologies can still provide satisfactory analog performance in terms of gain and power efficiency. This frequency band has been utilized in numerous works for wireless communication~\cite{Rebeiz1,Rebeiz2,Rebeiz3,Rebeiz4}. \lastrev{Recently D-band plastic waveguide links~\cite{reynaert1,reynaert2,reynaert3,reynaert,George2} have also emerged as very promising solutions for data-center communication links as a replacement for traditional copper wires especially for meter-range communications. Due to their potential they have recently been identified as one of the most promising emerging technologies~\cite{2026_trend}}.\\
\indent Antennas represent one of the fundamental key enabling components for D-band applications~\cite{antenna_survey}. Despite their crucial role, terahertz (THz) and sub-THz antennas face several technological challenges, most notably the high dielectric and conductor losses typically associated with conventional printed circuit board (PCB) materials at these frequencies, as well as the dimensional limitations inherent to PCB fabrication processes~\cite{antenna_survey}. Advanced technologies, such as organic-based packaging substrates (e.g. FCBGA, FCCSP or FOWLP) and low-temperature co-fired ceramic, can help address these issues~\cite{ceramic}. Many antenna configurations have been discussed for D-band applications, such as SIW antennas~\cite{SIW}, dipoles~\cite{dipole}, and patch antennas~\cite{George_Neelam}. However, in a recent survey~\cite{antenna_survey} of D-band antennas, no in-package design had successfully achieved a sufficiently wide bandwidth to cover the entire D-band. 
\\ \indent Dual-patch antennas are inherently wideband, due to their ability to support dual resonant modes, making them a good candidate for wideband D-band applications. \lastrev{In recent years, several dual-patch antennas have been proposed for wireless applications~\cite{payam,stacked_letter,stacked_third,stacked2,stacked_reviewer} as well as for dielectric waveguide applications~\cite{George_Neelam,George2}.} These designs rely on different technologies, such as flexible PCB, organic packaging and LTCC. These technologies offer complementary advantages: flex-PCB provides mechanical flexibility and low-cost manufacturing, while LTCC enables multilayer integration, precise dielectric control, and compatibility with millimeter-wave applications. However, these technologies often suffer from issues related to minimum allowable feature size and increased losses when compared to advanced organic packaging technologies. \lastrev{A common feature among these designs is the excitation of the lower patch from a slot-coupled grounded coplanar waveguide (GCPW). This feeding approach isolates the radiating element from the feeding one, and the planar nature of the GCPW facilitates compact integration of the antenna with flip-chip technologies, which is crucial for front-end D-band modules~\cite{payam,stacked2,stacked_reviewer}. Some designs, however, utilize transitions from the WR6 waveguide attached to the D-band extender to facilitate measurement and alignment as in~\cite{stacked_letter,stacked_third}.}
\begin{figure} [!t]
\centering%
\subfloat[]{%
\centering
\includegraphics[width=\linewidth]{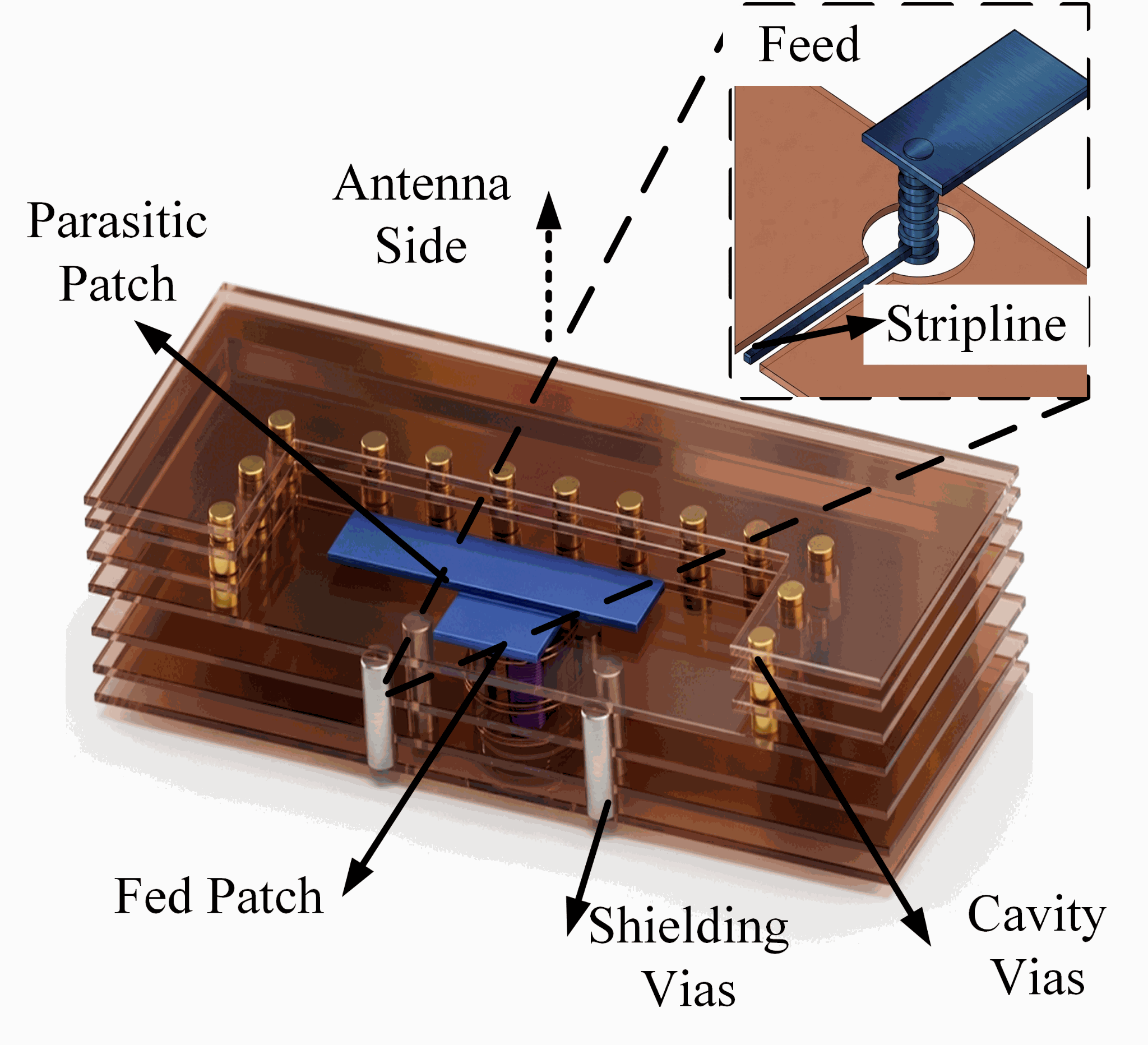}
\label{fig:1a}   
}%
\\[0.1 mm]
\subfloat[]{%
\centering
\includegraphics[width=\linewidth]{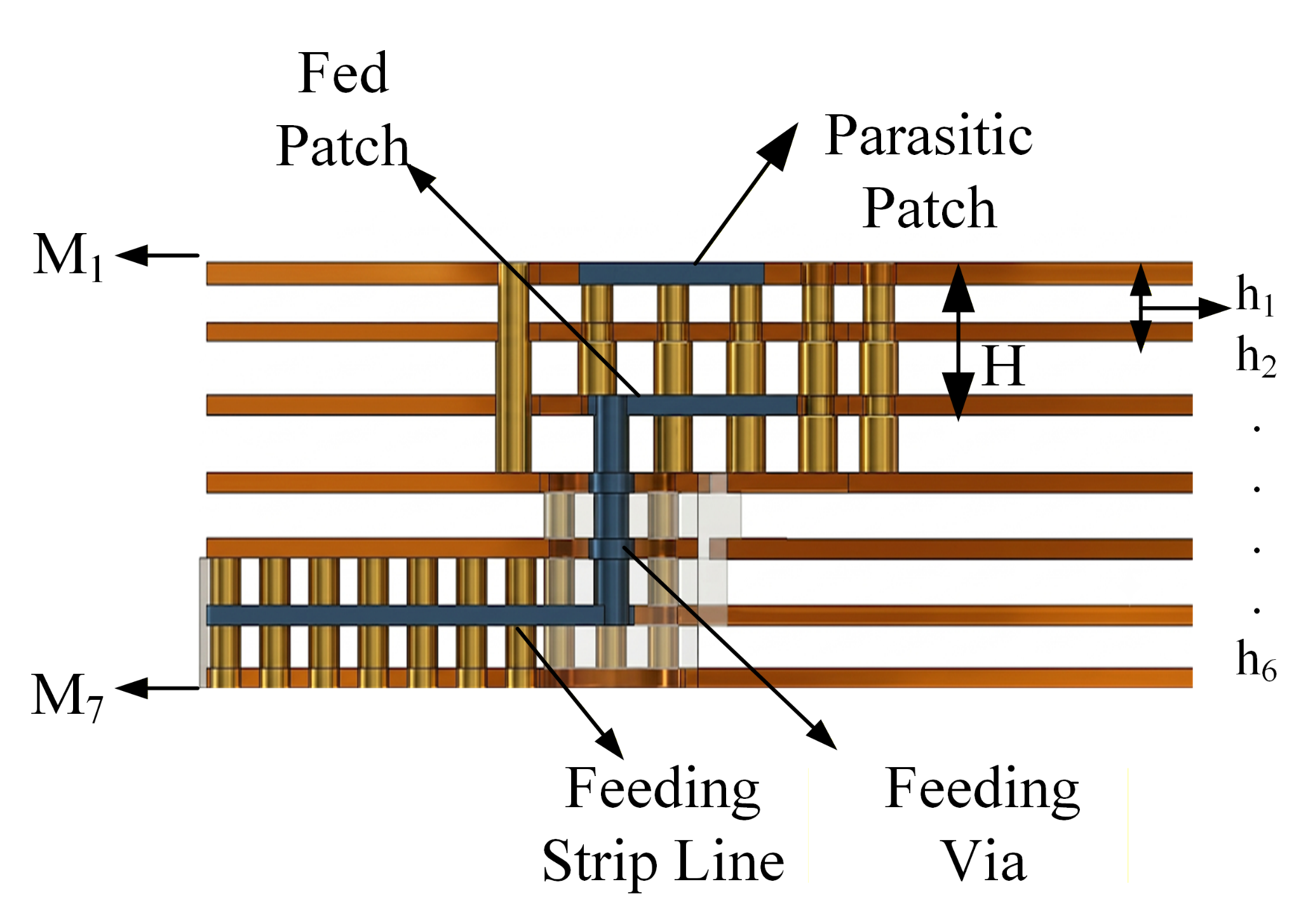}
\label{fig:1b}
}
\\[0.1 mm]
\subfloat[]{%
\centering
\includegraphics[width=\linewidth]{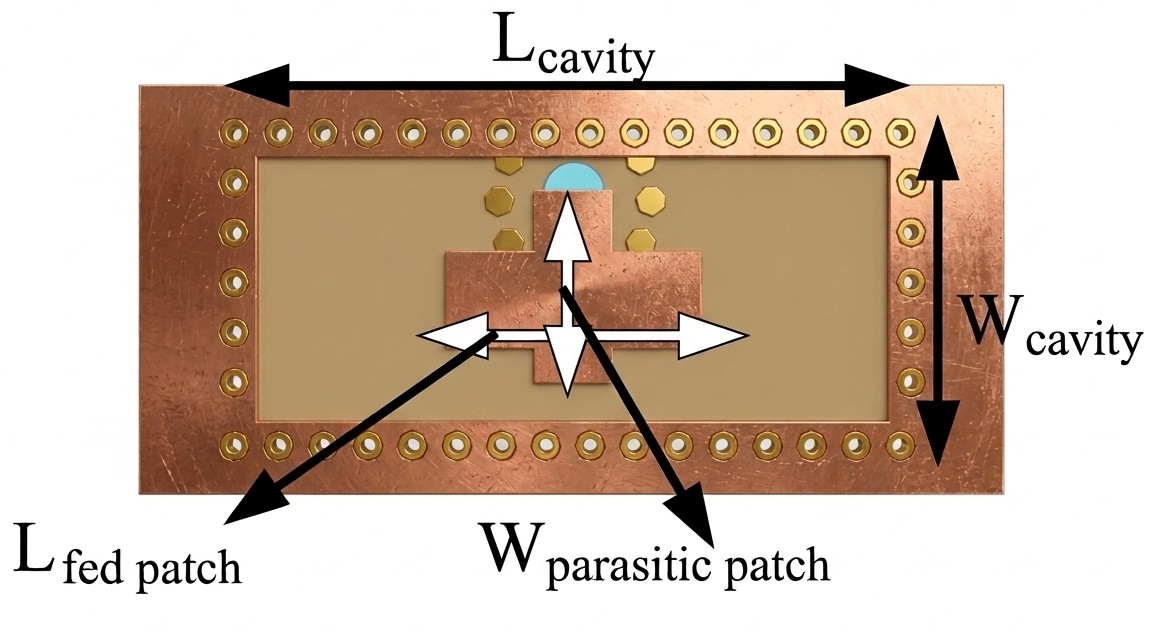}
\label{fig:1c}
}
\caption{\lastrev{Proposed structure of a strip-line fed stacked-patch antenna (a) Cut-away sectional view, (b) Side view of the Stackup (h is the height of the dielectric layers), (c) Top view showing the patches and their dimensions.}}
\label{fig:1}
\end{figure}
\lastrev{The previous designs showcased excellent performance and provided a design methodology for the two resonant modes of the stacked patch antennas. Most of them relied on a square type patch that is fed using a stacked via matched to a $50~\Omega$ line. Several techniques were proposed to enhance the impedance bandwidth, including the use of broadband transmission-line (TL) models, as reported in~\cite{payam}, as well as various open- and short-circuited stub-loading techniques~\cite{stacked_reviewer} applied to the excitation via to improve impedance matching. Despite these advancements, none of the reported designs achieved full coverage of the entire D-band $60~GHz$, with the widest reported impedance bandwidth for this class of antennas being approximately $40~GHz$. Furthermore, previous studies focused exclusively on either wireless communication or dielectric waveguide applications and did not explore antenna designs capable of operating efficiently in both free-space wireless links and dielectric waveguide interconnects. Such dual functionality is particularly important in the D-band, where both wireless communications and dielectric waveguide-based interconnects are emerging as key enabling technologies~\cite{2026_trend}. Moreover, this dual functionality is a very practical feature of the antenna when used in dielectric waveguide applications as it allows the independent experimental characterization of the IC transceivers using an over-the-air wireless channel.  This feature is of practical importance when setting up the dielectric waveguide transceiver as it enables decoupling the transceiver operation from any non-idealities of the waveguide channel itself. } \\ \indent
In this article, a wideband dual-patch antenna is proposed and fabricated on an organic packaging technology. \lastrev{The proposed antenna consists of a dual-stacked patch enclosed within a rectangular cavity formed by closely spaced metallized vias. Unlike a conventional stacked patch antenna, the cavity-backed configuration behaves as a coupled multi-resonator system in which the measured resonances correspond to the eigenmodes of the complete patch–cavity structure rather than to the individual patch resonances alone. These eigenmodes arise from the interaction between the lower and upper patch modes and one or more cavity modes supported by the enclosure~\cite{patch_numerical}. Electromagnetic coupling between the stacked patches gives rise to the conventional in-phase and out-of-phase hybrid modes, while additional coupling with the cavity modes produces hybrid patch–cavity resonances. When a cavity resonance lies in close spectral proximity to a patch resonance, mode hybridization occurs, resulting in frequency splitting and the addition of two more resonances. The strength of this interaction is controlled by the vertical spacing between the stacked patches, which simultaneously influences the cavity resonance and the patch–cavity coupling. 
\\ \indent To realize broadband operation, the antenna is designed to maintain relatively weak direct coupling between the patches through an appropriate vertical separation, while the rectangular aspect ratio of the patches is chosen to align the dominant patch modes with the cavity modes, thereby maximizing their interaction. This engineered mode hybridization produces four hybrid resonant modes that merge to form a wide impedance bandwidth, enabling a full coverage of the entire D-band. The proposed broadband matching technique extends the applicability of the antenna beyond conventional wireless communications. Specifically, the antenna also serves as an efficient coupler to a single-mode dielectric waveguide, achieving full D-band operation with low insertion loss. To the best of the authors' knowledge, this is the first reported antenna design capable of operating efficiently in both wireless communication and dielectric waveguide interconnect applications. The feeding patch is fed using a stacked-vertical via, which can be fed using a stripline for integration with a flip-chip packaging technology. Additionally, for direct characterization using a VNA, a transition antenna is also proposed in this work for direct-feeding via a WR6 waveguide eliminating any need of a probe-feeding transition.}\\
\indent The antenna parameters are experimentally characterized \lastrev{for wireless applications} in an anechoic chamber using a $20~dBi$ D-band horn antenna as a reference and electronic beam alignment, demonstrating superior bandwidth and radiation efficiency. \lastrev{The antenna is also characterized for dielectric waveguide excitation using the depressed-core dielectric waveguide proposed in~\cite{MTT_mine} showing superior performance and a return loss that covers the entire D-band. The remainder of this paper is organized as follows. Section II presents the antenna design methodology, with particular emphasis on the bandwidth extension using mode hybridization of the stacked patches within the cavity. Section III discusses the design and the concurrent impedance-matching considerations that enable efficient operation for both wireless communication and dielectric waveguide interconnect applications. Section IV presents the experimental validation and measured results. Finally, Section V concludes the paper.} 
\begin{figure}[!t]
\subfloat[]{%
\centering
\includegraphics[width=\linewidth]{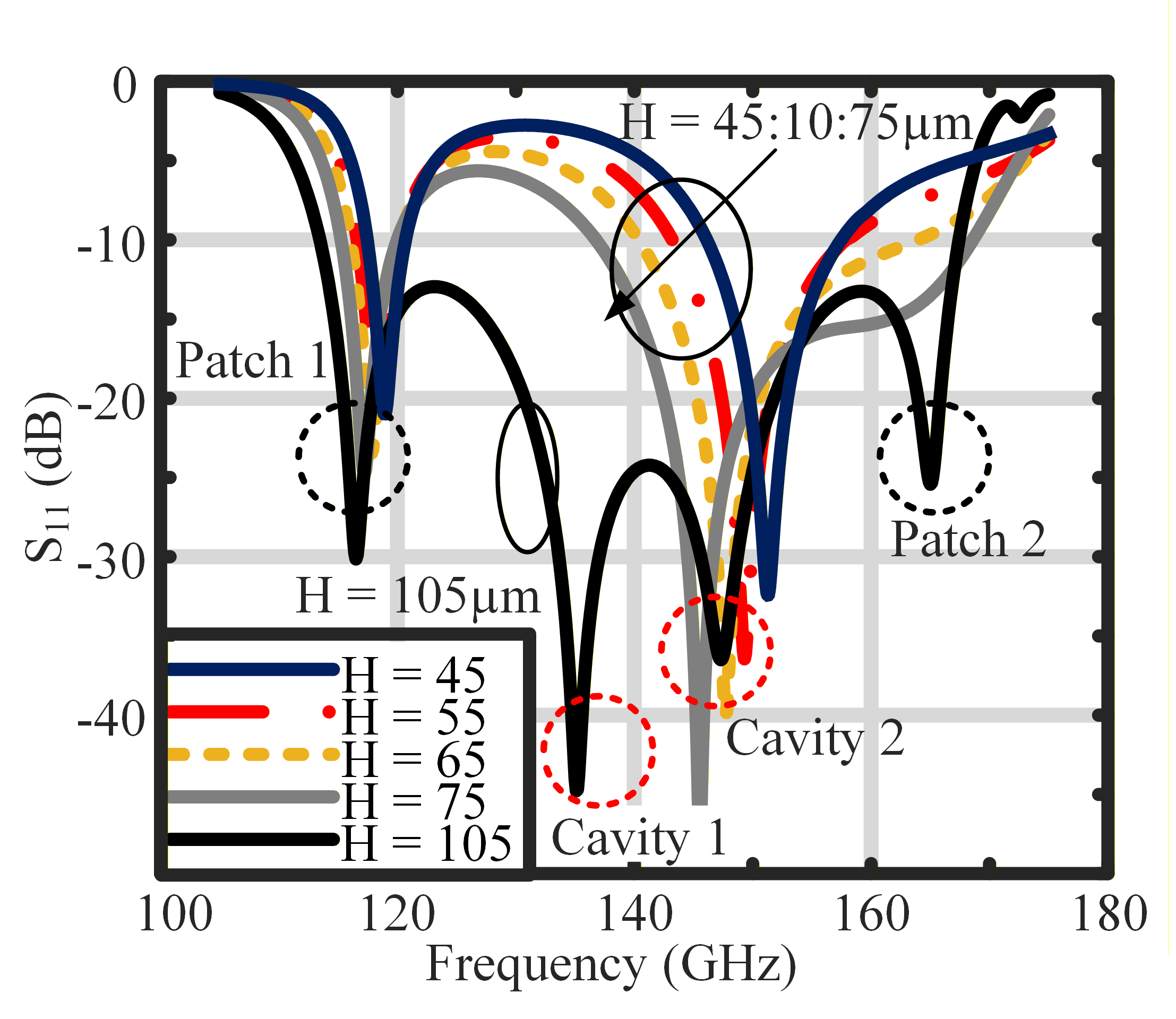}
\label{fig:resonances_height}   
}%
\\[0.1 mm]
\subfloat[]{%
\centering
\includegraphics[width=\linewidth]{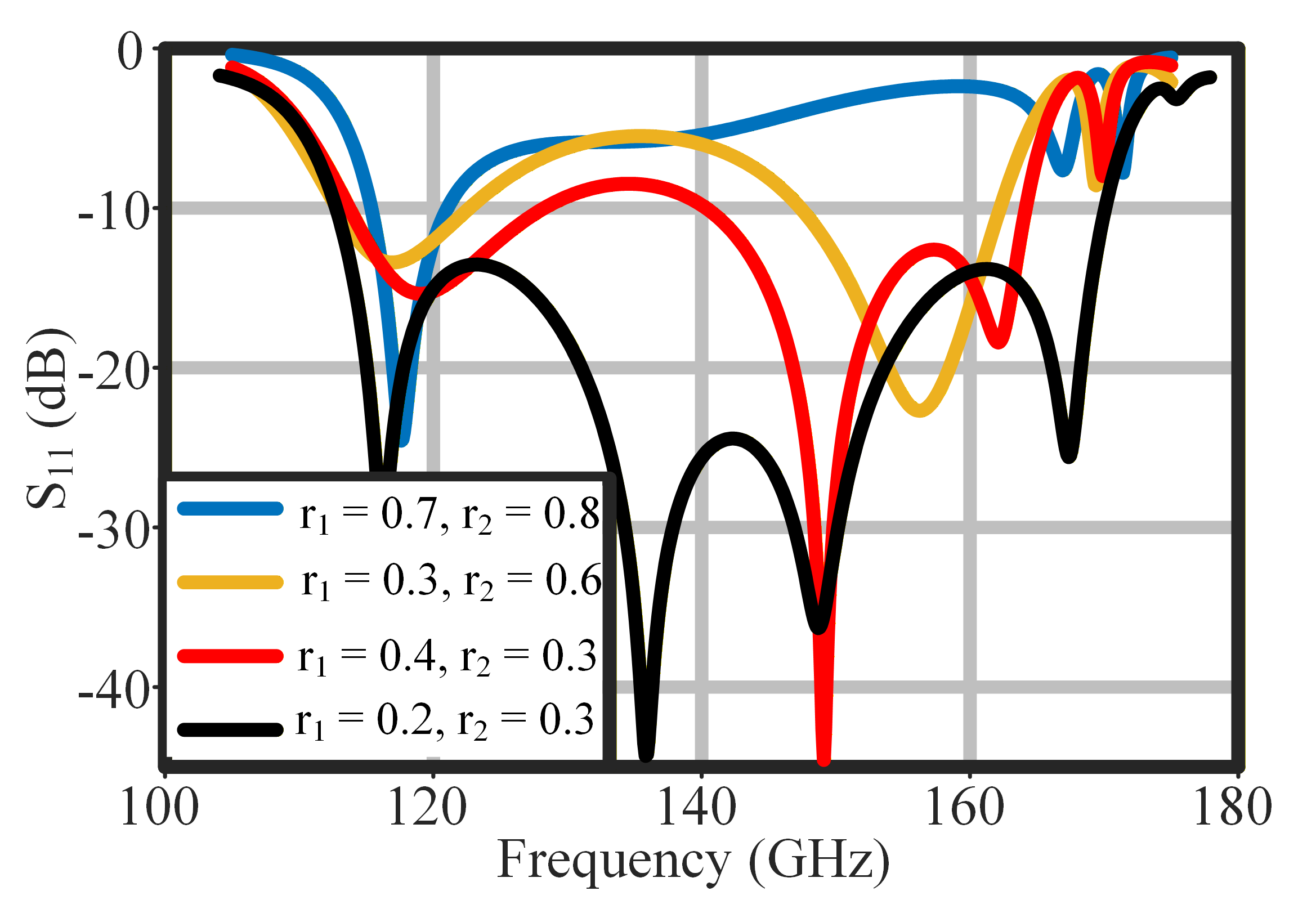}
\label{fig:aspect_resonances}
}
\caption{\lastrev{Effect of the:  (a) inter-patch vertical spacing $H$, (b) patch aspect ratio $r_1$ and $r_2$ at $H = 105~\mu m$ on the in-band resonances of a cavity-enclosed dual-patch antenna.}}
\label{fig:resonances_height_patches}
\end{figure}
\begin{figure}
\centering%
\subfloat[]{%
\centering
\includegraphics[width=0.8\linewidth]{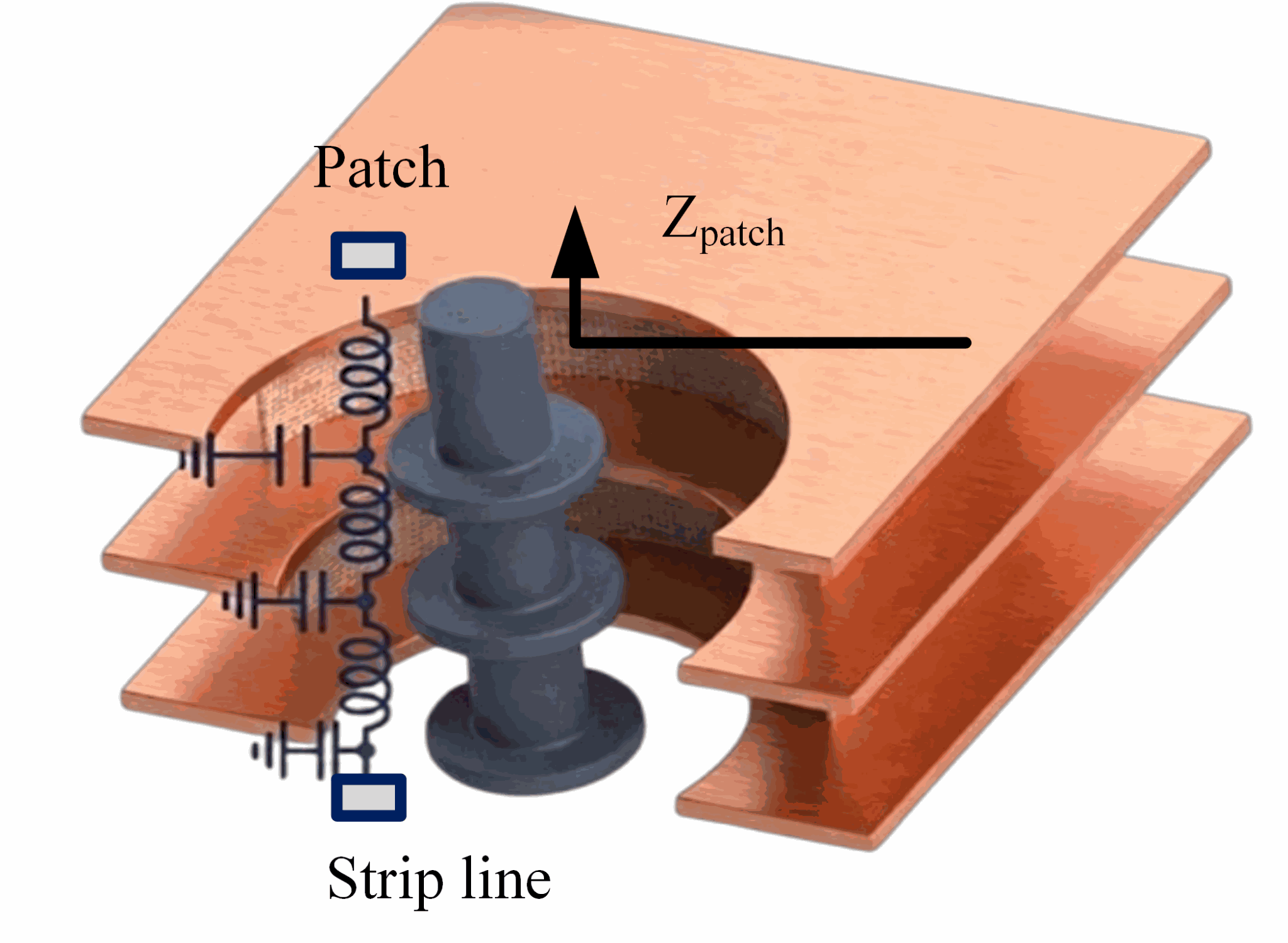}
\label{fig:strip_via}   
}%
\\[0.1 mm]
\subfloat[]{%
\centering
\includegraphics[width=\linewidth]{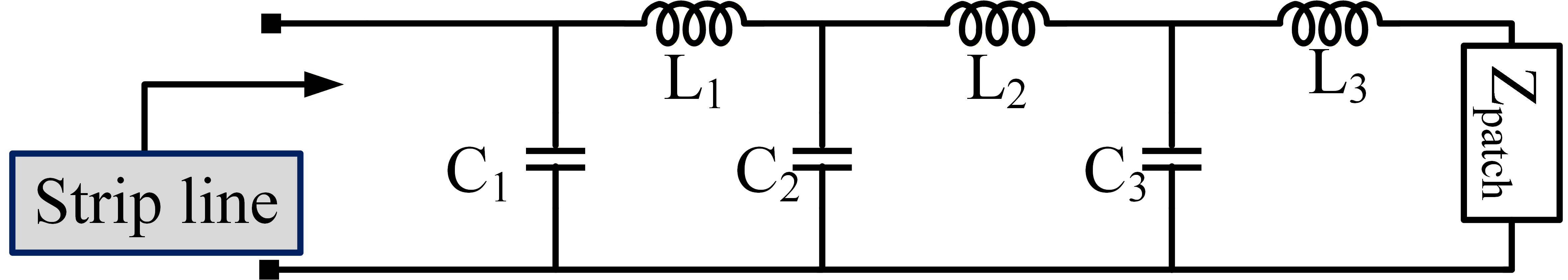}
\label{fig:strip_via_matching}
}
\caption{\lastrev{The feeding via of the dual-patch antenna (a) Schematic, (b) Equivalent circuit models.}}
\label{fig:matching_considerations}
\end{figure}
\section{\lastrev{Cavity-enclosed Dual Patch System}}
\lastrev{The proposed dual-patch antenna is depicted in Fig.~\ref{fig:1}. The antenna is integrated within a via-defined cavity and is excited through a vertical feeding via, as illustrated in Fig.~\ref{fig:1a}. The excitation signal is delivered via a stripline that can be routed to a flip-chip interconnect on the organic substrate. The antenna is implemented using a seven-layer organic package technology, as shown in Fig.~\ref{fig:1b}, and leverages mode hybridization between the dual-patch structure and the resonant modes of the enclosing cavity. This interaction creates four resonances within the operating band, resulting in an enhanced impedance bandwidth. Consequently, the antenna achieves a $10~dB$ return-loss bandwidth that fully covers the entire D-band spectrum.}\\
\indent \lastrev{To achieve the desired mode hybridization and generate the required in-band resonances, several key design parameters must be carefully optimized, as illustrated in Fig.~\ref{fig:1c}. These parameters include the cavity dimensions (length and width), the dimensions of the fed and parasitic patches ($L$ and $W$), and the inter-patch spacing ($H$). The proposed dual-patch structure inherently supports two primary resonant modes, which are mainly determined by the dimensions of the fed and parasitic patches, respectively. As a first-order approximation, the resonant frequencies of the fed and parasitic patches can be estimated from the dimensions of their respective cavities using the preceding formulations~\cite{payam}. Unlike previous works~\cite{payam,stacked_reviewer}, which utilized square patches, this work adopts rectangular patches to provide additional degrees of freedom for controlling the coupling between the patch and cavity modes and promoting effective mode hybridization. Accordingly, orthogonal patch modes are realized by designing the fed patch such that $L_{fedpatch} > W_{fedpatch}$ and the parasitic patch such that $W_{parasiticpatch} > L_{parasiticpatch}$. This makes $L_{fedpatch}$ and $W_{parasiticpatch}$ resposible for the in-band resonance frequencies of the fed and parasitic patches, respectively, since they correspond to the longest dimension of each. Thus, from~\cite{payam}, to set the in-band resonance of the fed patch, its length is set to:}
\begin{equation}
L_{fedpatch} = \frac{c}{2f_o\sqrt{\frac{\varepsilon_r+1}{2}}}
\label{eqn:1}
\end{equation}
\lastrev{The width of the fed patch, which is smaller than its length, can be defined based on a new parameter determined by its ratio to the length as given in:
\begin{equation}    
r_1 = W_{\mathrm{fedpatch}}/L_{\mathrm{fedpatch}}
\label{eqn:r_1}
\end{equation}
where $0\leq r_1 \leq 1$, and an initial value of 0.5 is selected.
On the other hand, to set the in-band resonance of the parasitic patch, since the width is its longest dimension, we choose it to be:}
\begin{equation}
\begin{split}
W_{parasitic patch} = \frac{c}{2f_o\sqrt{\varepsilon_{eff}}} - \\0.842 H (\frac{(\varepsilon_{eff} + 0.3)(\frac{W_{fedpatch}}{H}+0.264)}{(\varepsilon_{eff} -0.258)(\frac{W_{fedpatch}}{H}+0.8)})
\end{split}
\label{eqn:2}
\end{equation}
\lastrev{Where $\varepsilon_{eff}$ is the effective dielectric constant of the microstrip patches and $H$ is the height or spacing between the two patches. The parasitic patch length, which is smaller than its width, can also be defined as its ratio to the width as given in (\ref{eqn:r_2}). where $0 \leq r_2\leq 1$ and an initial value of 0.5 is selected.
\begin{equation}
    r_2 = L_{\mathrm{parasiticpatch}}/W_{\mathrm{parasiticpatch}}
    \label{eqn:r_2}
\end{equation}
Note that $r_1$ is defined as width over length and $r_2$ is defined as length over width such that both ratios are between 0 and 1 due to the parasitic patch being tall and the fed patch being wide. The cavity dimensions are then determined according to the resonance frequency expression~\cite{pozar}}.
\begin{equation}
f_{mn} = \frac{c}{2\sqrt{\varepsilon_o\varepsilon_r \mu_o}} \sqrt{(\frac{m}{W})^2 + (\frac{n}{L})^2}
\label{eqn:3}
\end{equation}
\lastrev{For these modes to generate the required resonances inside the D-band $110-170~GHz$, the width and the length of the cavity $W_{cavity}$ and $L_{cavity}$ are chosen such that the mode resonances are orthogonally directed along the X and Y directions i.e. $m = 0$ and $n = 0$ for the two modes respectively.} 
\\ \lastrev{\indent To illustrate the impact of inter-patch spacing and patch geometry on mode hybridization, the dual-patch resonances were initially designed to occur near $110~GHz$ and $170~GHz$, respectively, using (\ref{eqn:1}) and (\ref{eqn:2}). The cavity dimensions were subsequently selected such that its dominant resonant modes occur at approximately $130~GHz$ and $150~GHz$. The inter-patch spacing ($H$) was then swept, and the resulting resonance behavior is presented in Fig.~\ref{fig:resonances_height}. \\
\indent \lastrev{The figure shows that when the spacing between the patches is small, resulting in strong electromagnetic coupling, only two resonances are observed within the operating band, as illustrated by the first four curves. This behavior is consistent with previously reported stacked dual-patch antenna designs~\cite{payam,stacked_reviewer}, in which strong coupling between the resonators gives rise to two hybridized modes. As the inter-patch spacing increases, the coupling strength decreases, allowing additional resonant modes to emerge and become distinguishable within the operating band.} In contrast, increasing the inter-patch spacing reduces the mutual coupling between the patches and promotes stronger interaction between the patch and cavity modes. Consequently, when the spacing exceeds approximately $100~\mu$m, effective mode hybridization occurs, giving rise to two additional in-band resonances. This mechanism enables the realization of four resonances within the desired frequency range, thereby significantly enhancing the impedance bandwidth. \\
\indent Better patch-cavity mode hybridization also requires the patch modes to be aligned with the cavity modes. This requires that $r_1$ (\ref{eqn:r_1}) and $r_2$ (\ref{eqn:r_2}) to be chosen such that the patch modes align better with the cavity modes for wider bandwidth operation. Having established the spacing requirements for efficient mode hybridization, the aspect ratios must be optimized to maximize the impedance bandwidth.
To quantify this influence, the aspect ratios of the fed and parasitic patches $r_1$ and $r_2$ were swept independently, and the corresponding results are shown in Fig.~\ref{fig:aspect_resonances}. The optimum impedance matching performance was achieved when both aspect ratios were maintained within the range of approximately 0.2 to 0.3, indicating that elongated rectangular patches provide the most favorable conditions for mode hybridization and broadband operation.}
\begin{figure}[!t]
\centering
\subfloat[]{%
\centering
\includegraphics[width=0.8\linewidth]{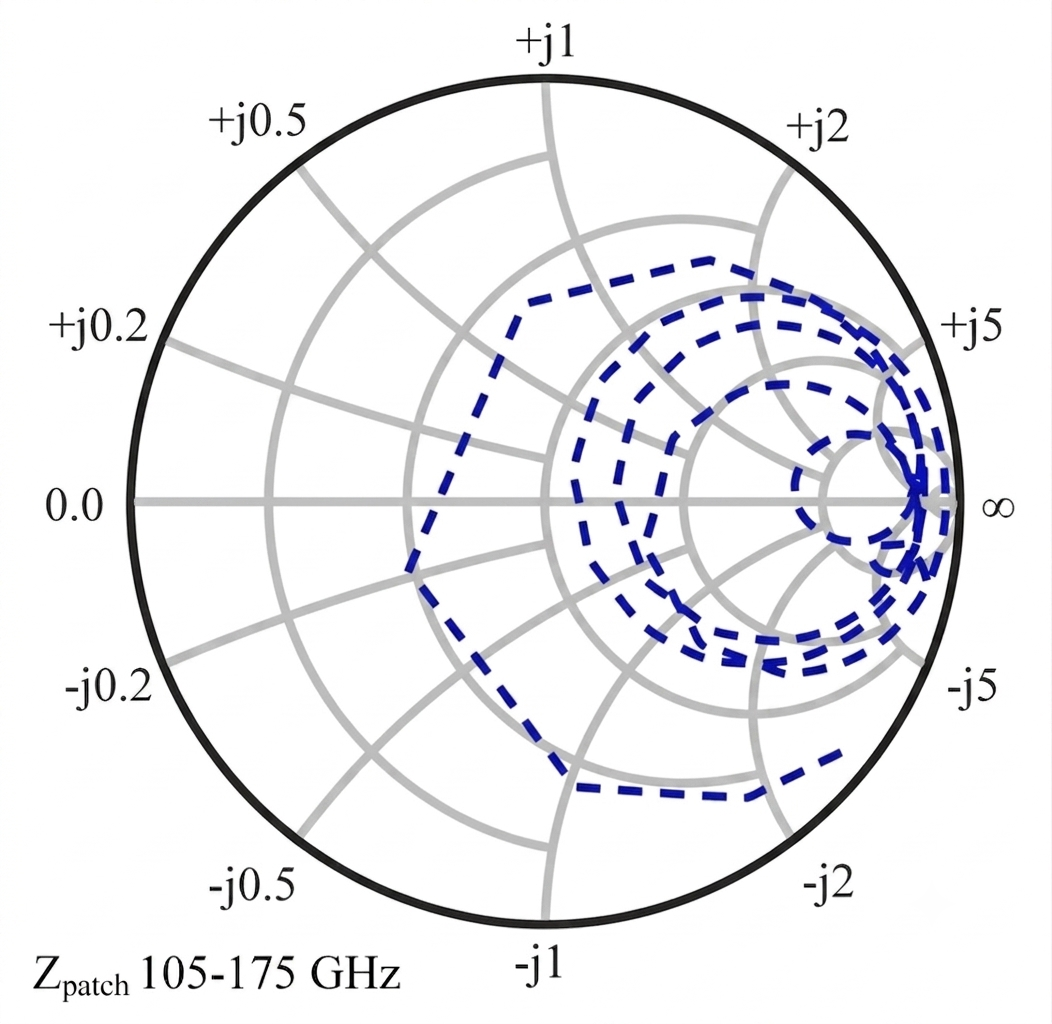}
\label{fig:smith_matching_wireless}   
}%
\\[0.1 mm]
\subfloat[]{%
\centering
\includegraphics[width=\linewidth]{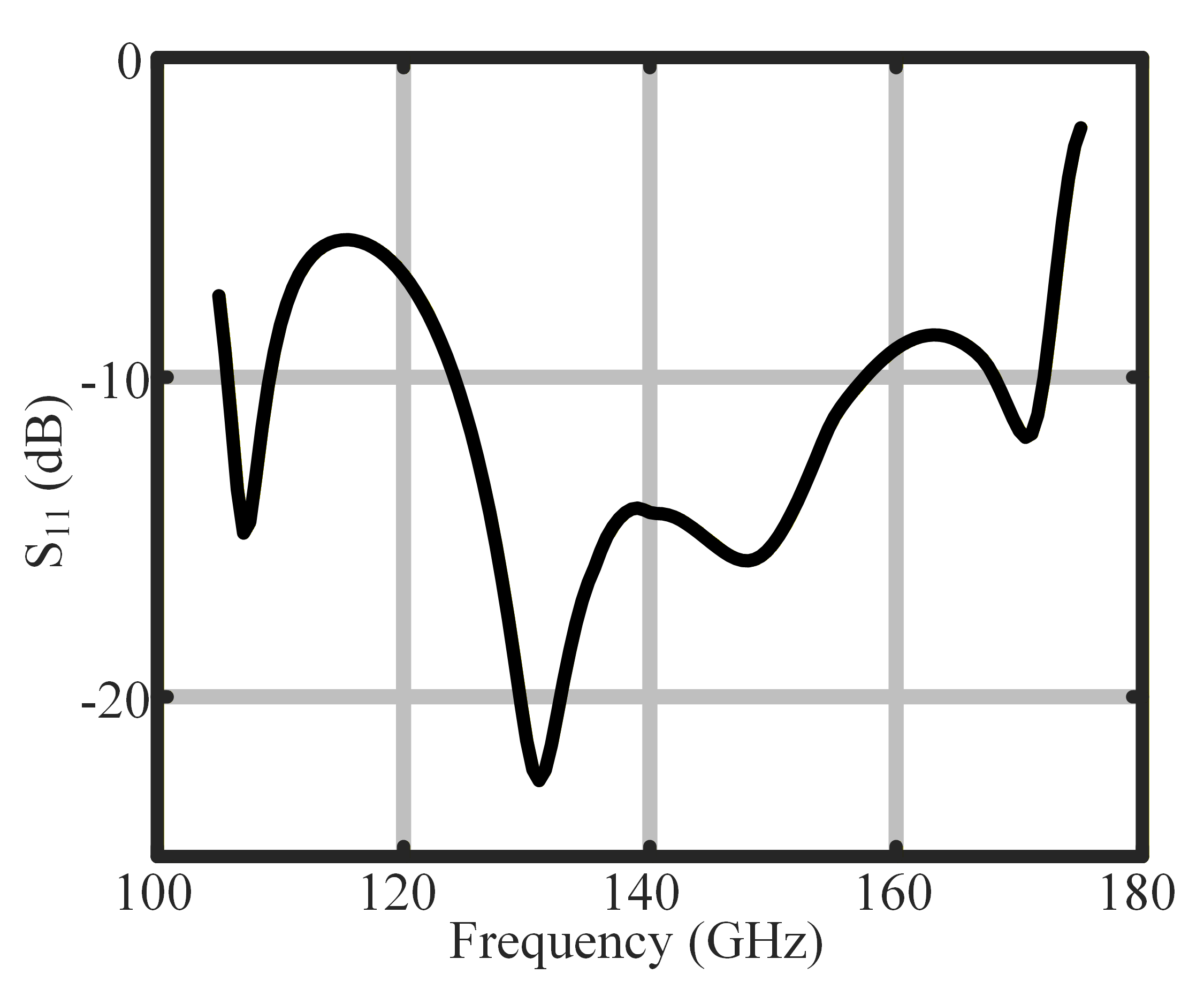}
\label{fig:smith_matching_RL}
}
\caption{\lastrev{Wireless matching consideration (a) The input impedance $Z_{patch}$ required for matching, (b) RL at the starting point}}
\label{fig:matching_WG}
\end{figure}
\begin{figure}[!t]
\centering
\subfloat[]{%
\centering
\includegraphics[width=\linewidth]{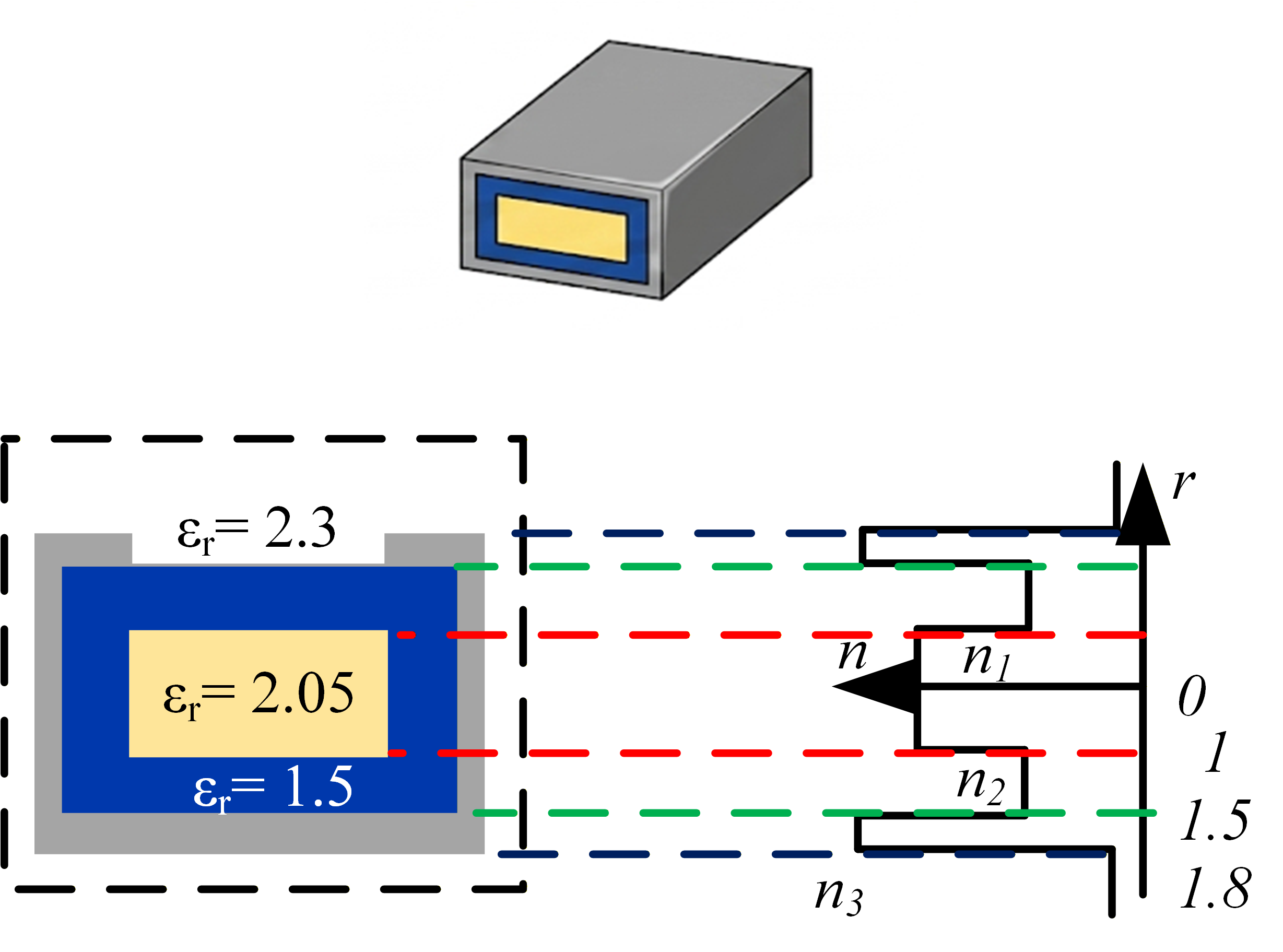}
\label{fig:waveguide}   
}%
\\[0.1 mm]
\subfloat[]{%
\centering
\includegraphics[width=0.8\linewidth]{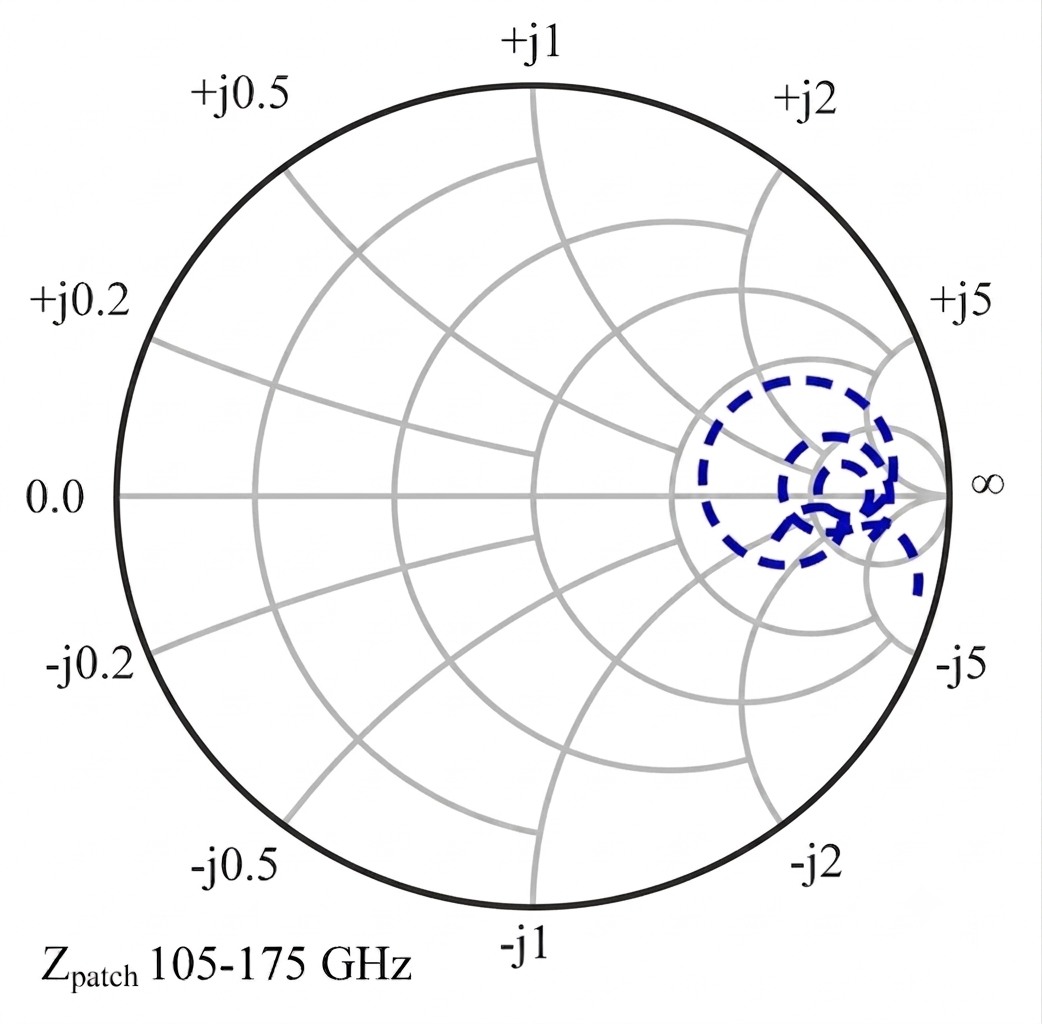}
\label{fig:WG_smith_matching}
}
\caption{\lastrev{Waveguide matching consideration (a) M-step waveguide structure, (b) $Z_{patch}$.}}
\label{fig:matching_waveguide}
\end{figure}
\section{\lastrev{Matching Consideration and Dielectric-Waveguide Excitation}}
\lastrev{The proposed antenna is fed using a $50~\Omega$ stripline that is connected to the feeding patch using a stacked-via transition, as illustrated in Fig.~\ref{fig:strip_via}. This transition can be represented by the equivalent circuit shown in Fig.~\ref{fig:strip_via_matching}. The via predominantly exhibits inductive behavior, while the via pads and their associated ground reference structure introduce a shunt capacitance. Consequently, the matching network is primarily realized through appropriate selection of the pad and antipad dimensions. \\
\indent The equivalent circuit model, shown in Fig.~\ref{fig:strip_via_matching}, provides useful initial design guidelines, where the inductive and capacitive parasitics can be extracted from standalone full-wave simulations of the via transition. However, accurate impedance matching ultimately requires full-wave electromagnetic (EM) optimization of the pad and antipad geometries, as well as the via feeding location relative to the center of the feeding patch. This optimization procedure must be carried out for both the wireless-radiation and dielectric-waveguide excitation cases.\\
\indent This section discusses the matching considerations for the two configurations and presents the selected common design point, which achieves excellent matching for both. To accelerate the optimization process, an simplified model employing a fully metallized trench cavity, without sidewall vias, can first be analyzed. In this simplified setup, the antenna is excited by a 50 $\Omega$ stripline through a via transition, while the connection between the stripline and the lower patch is implemented using the minimum pad and antipad dimensions permitted by the packaging technology. This approach enables rapid evaluation of the matching performance prior to detailed optimization and finetuning of the complete structure.}
\begin{figure}[!t]
\centering
\includegraphics[width=\linewidth]{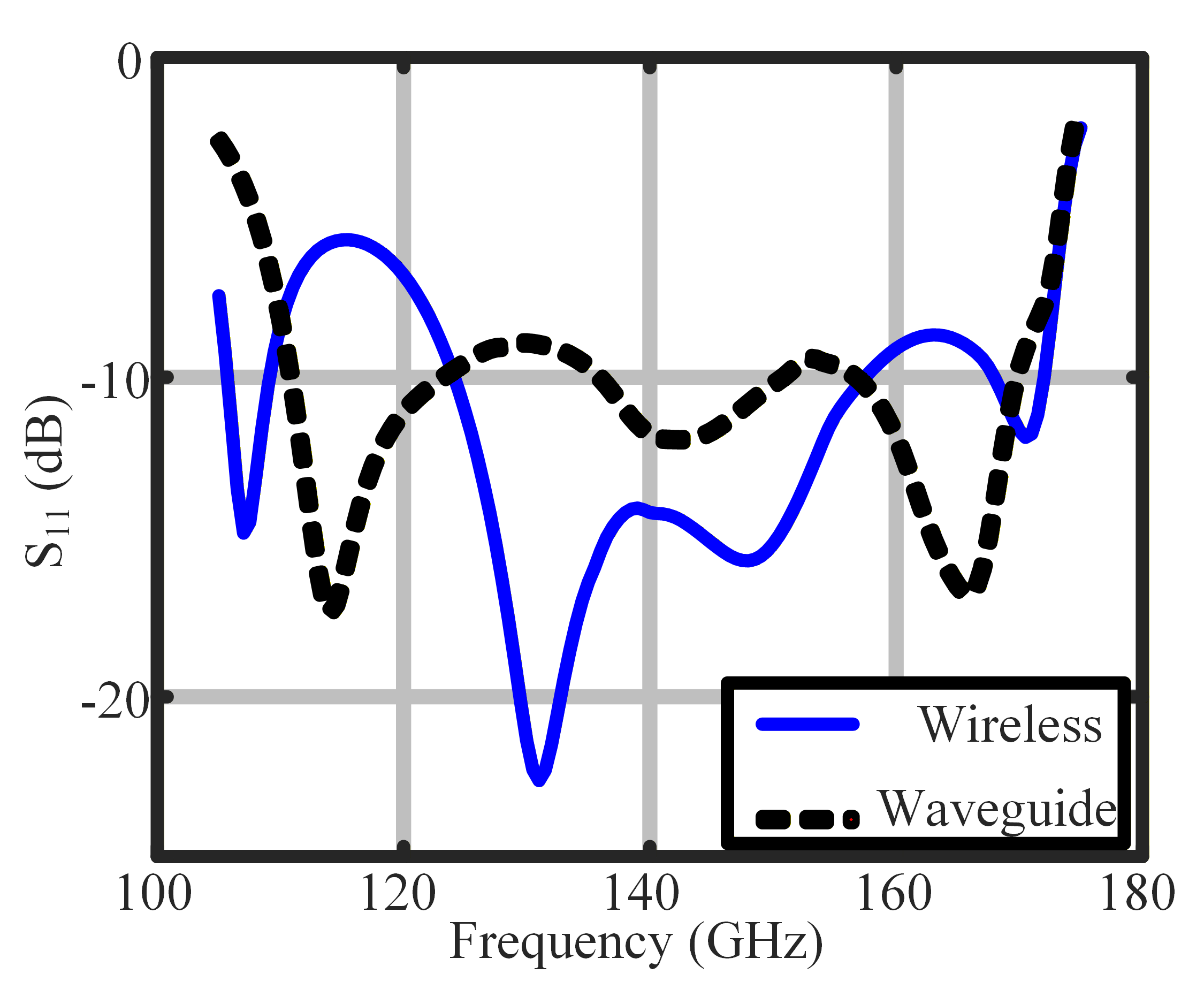}
\caption{\lastrev{The return loss of both the wireless case and the waveguide case in the initial design point.}}
\label{fig:wireless_WG_RL}
\end{figure}
\begin{figure}[!t]
\centering
\subfloat[]{%
\centering
\includegraphics[width=\linewidth]{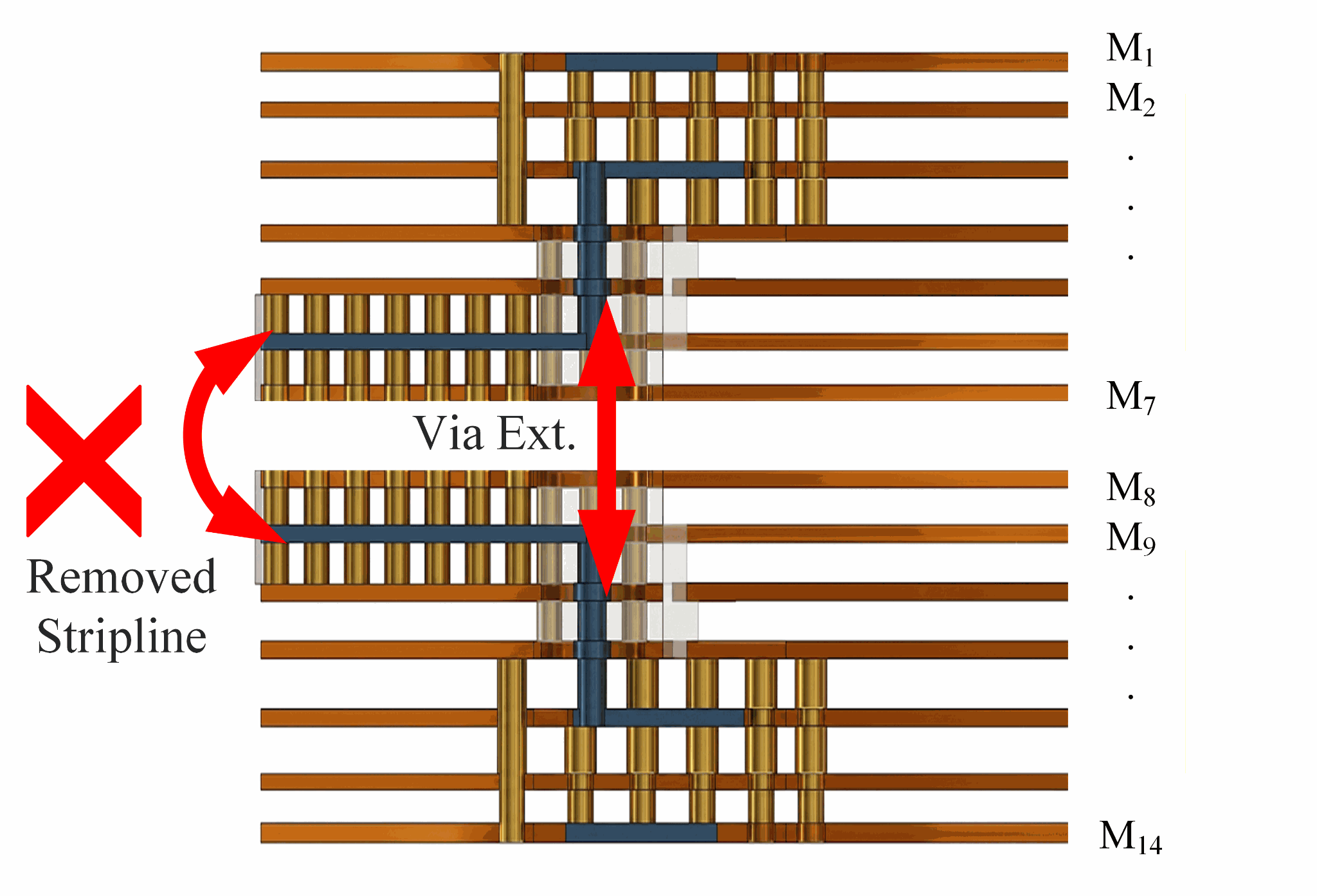}
\label{Fig:structure_building}   
}%
\\[0.1 mm]
\subfloat[]{%
\centering
\includegraphics[width=\linewidth]{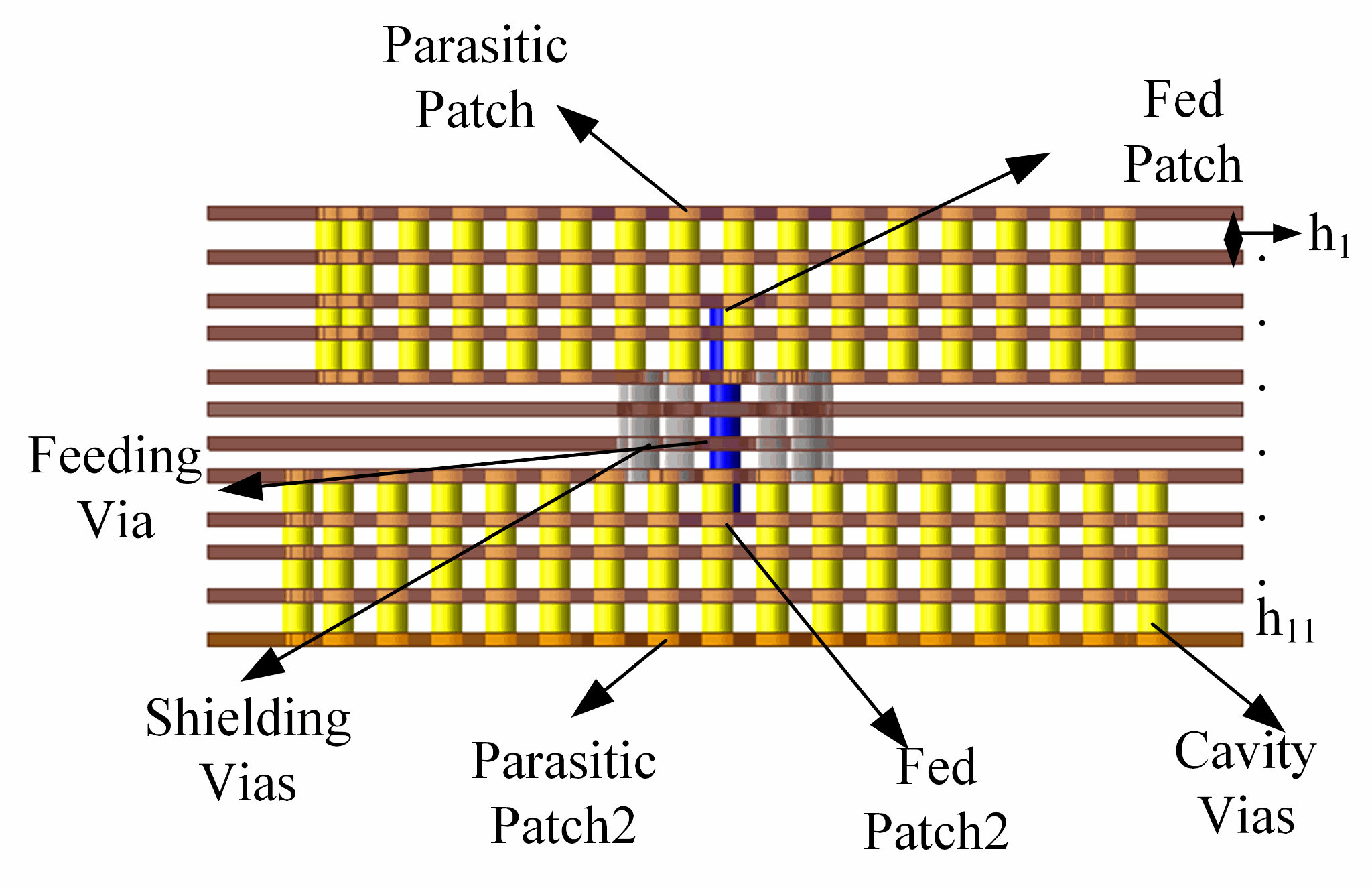}
\label{fig:full_structure_stackup}
}
\\[0.1 mm]
\centering
\subfloat[]{%
\centering
\includegraphics[width=\linewidth]{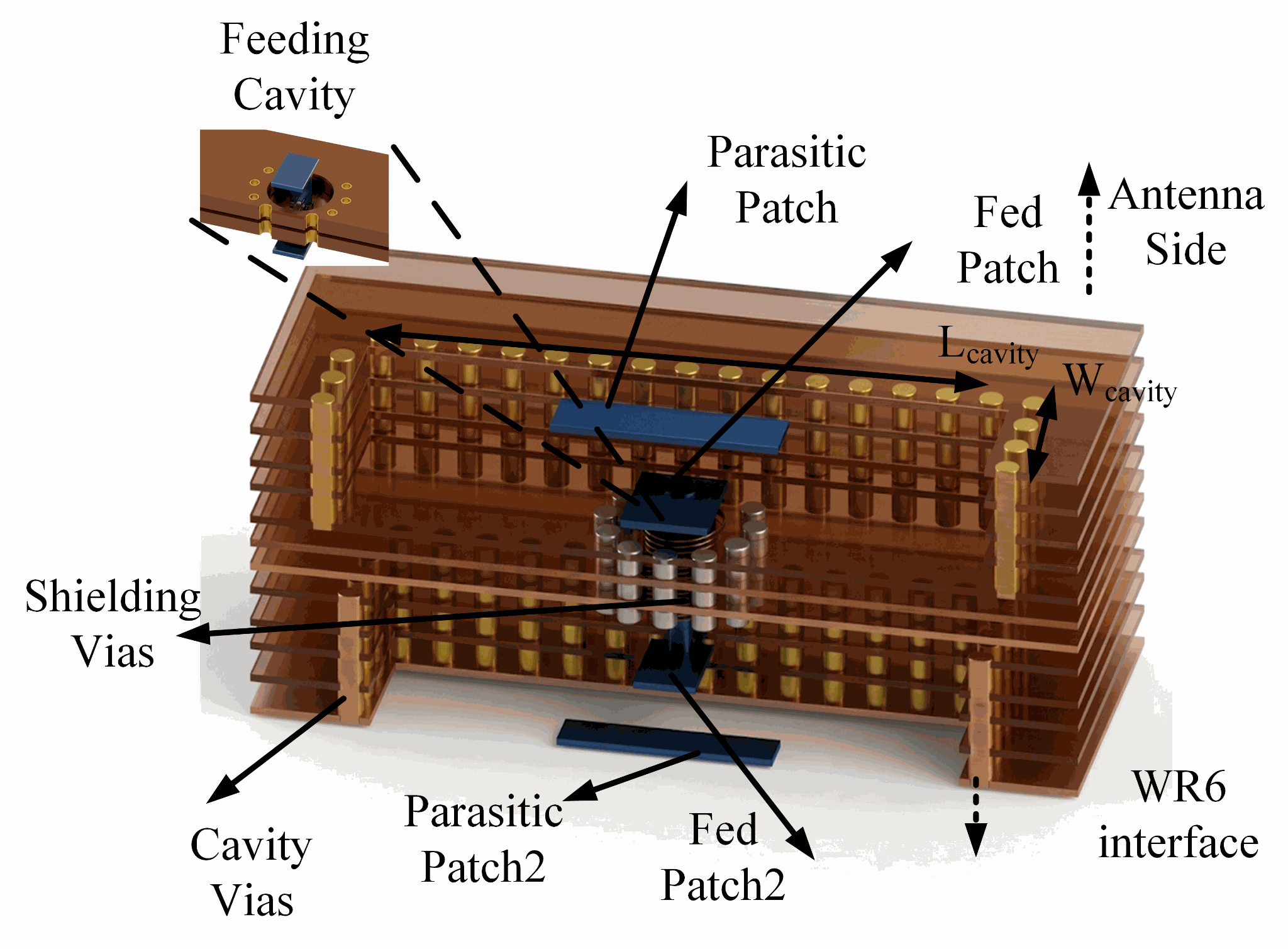}
\label{fig:full_structure_cut-out}   
}%

\caption{\lastrev{Proposed back-to-back stacked-patch antenna structure. (a) Structural integration of the two back-to-back antenna sections into a unified twelve-layer transition, illustrating the removal of the stripline layers ($M6$ and $M9$), merging of the adjacent ground planes, and extension of the feeding vias to form a continuous electromagnetic path. (b) Cross-sectional stackup of the resulting twelve-layer back-to-back stacked-patch structure (h is the height of the dielectric layers), (c) Cutaway view of the integrated structure, highlighting the four stacked patches. The WR6 launcher excites the D-band antenna through a shielded stacked via, highlighted in blue ($L_{Cavity}$ and $W_{Cavity}$ are the length and the width of the cavity given in Table~\ref{table1}.}}
\label{fig:full_structure}
\end{figure}
\begin{figure}[!t]
\centering
\subfloat[]{%
\centering
\includegraphics[width=\linewidth]{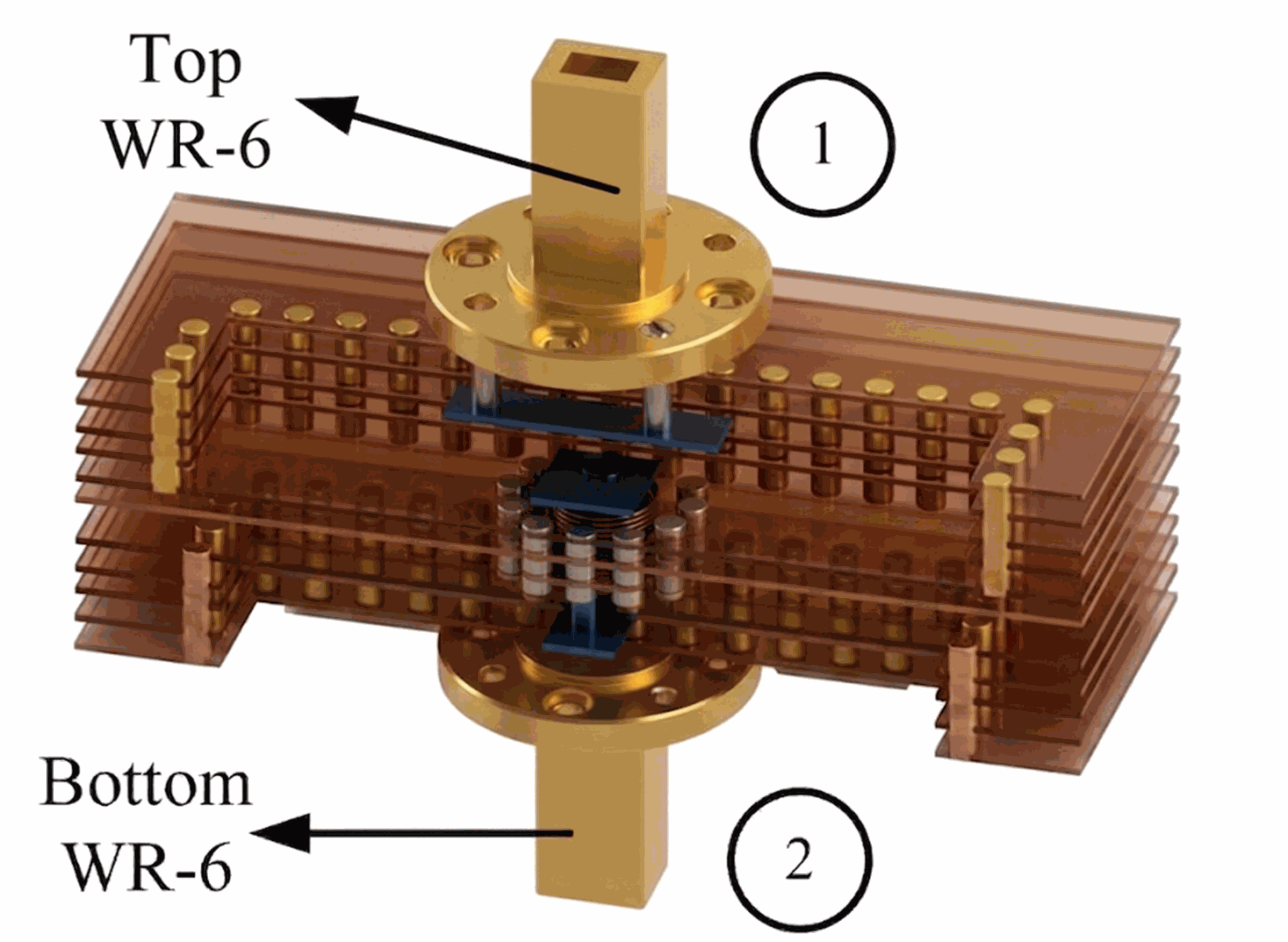}
\label{fig:WR6_back_to_back}   
}%
\\[0.1 mm]
\subfloat[]{%
\centering
\includegraphics[width=\linewidth]{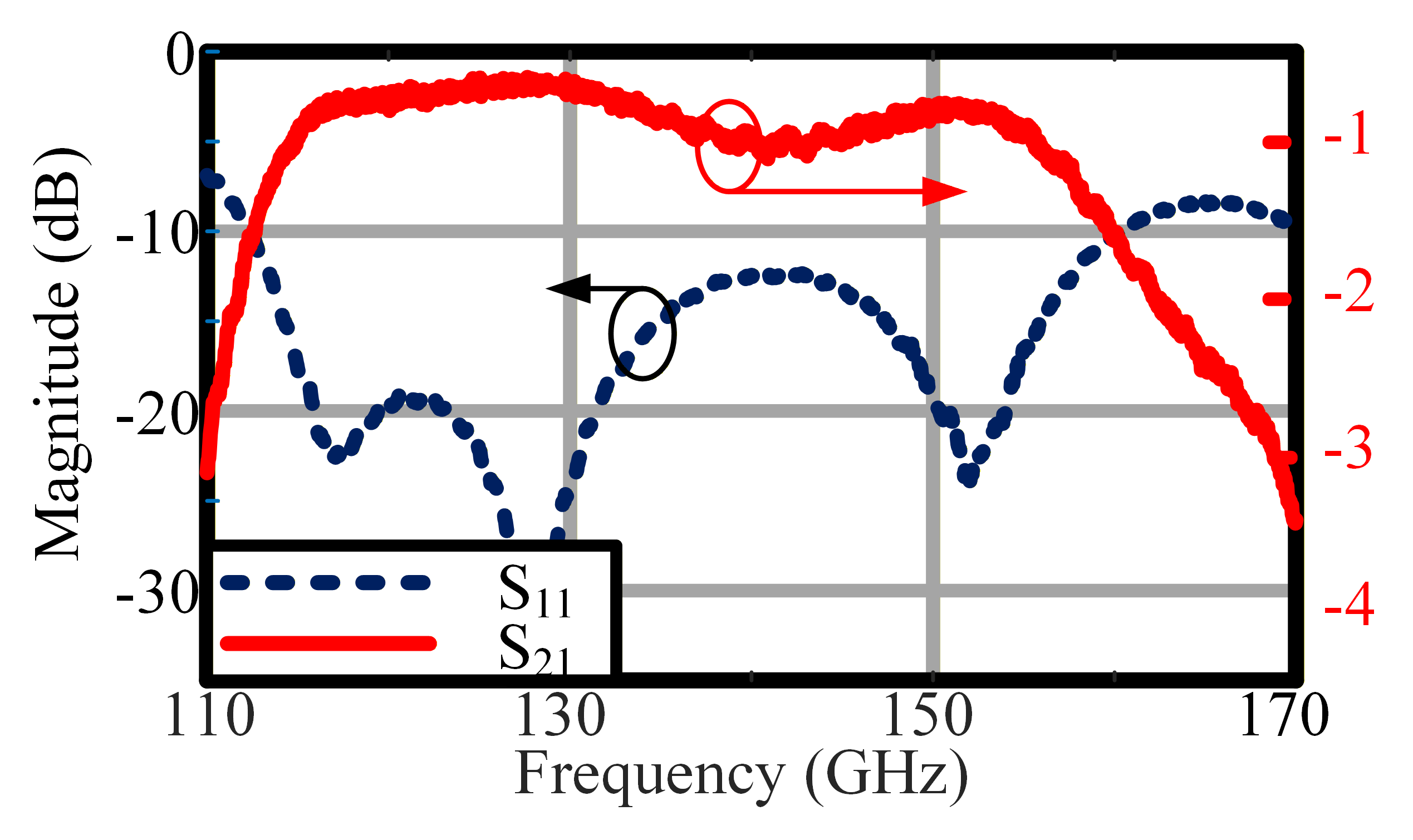}
\label{fig:transition}
}
\caption{\lastrev{Proposed structure of a back-to-back WR-6 to WR-6 transition for de-embedding and performance evaluation (a) Schematic. Note that the WR6 flanges whose dimensions are  ($L = 1.651~mm$ and $W=0.825~mm$) are not in scale to improve visibility the real structure dimensions are given in Table~\ref{table1}, (b) Measured Return and Insertion Loss}}
\label{fig:back_to_back}
\end{figure}
\begin{figure}[!t]
\centering
\includegraphics[width=\linewidth]{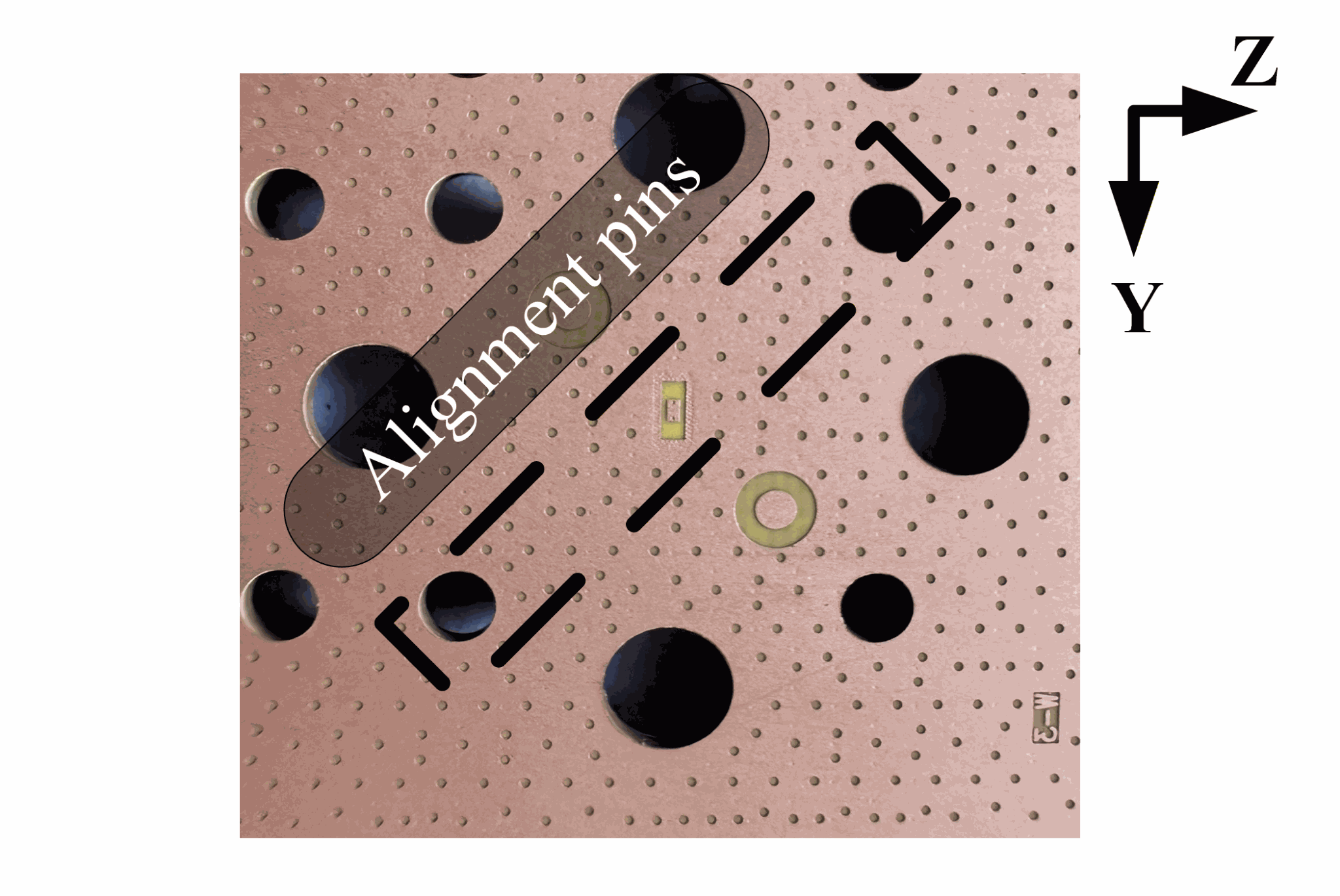}
\caption{The fabricated antenna on the twelve-layer organic package. Note the through holes are designed to match the alignment pins and screws of a WR-6 flange.}
\label{fig:fab_antenna}
\end{figure}
\subsection{Matching Considerations}
\lastrev{Matching the proposed antenna over a wide bandwidth has been made possible using the mode hybridization technique described in the previous section. The antenna has to be matched to the input impedance illustrated on Smith chart as shown in Fig.~\ref{fig:smith_matching_wireless}. The matching is achieved first by choosing the cavity and patch dimensions to give four resonances in-band. Following this, the focus shifts to the design of the feeding transition, including the optimization of the via pad and antipad dimensions as well as the heights of the constituent layers to achieve the desired impedance matching performance. A starting point is established by choosing a patch spacing $H$ of $105~\mu m$, i.e., $h_1 = h_2 = 60~\mu m$ and $t_{metal} = 15~\mu m$ with all the other dielectric layer heights set to $40~\mu m$. The via pads are assumed to be $70\mu m$ in diameter while the antipad diameters are chosen to be $100~\mu m$.  This configuration serves as the initial design point for the subsequent optimization process. As can be observed, it already exhibits favorable broadband impedance matching characteristics, with the impedance trajectory remaining relatively close to the center of the Smith chart throughout the frequency band of interest. The corresponding return-loss response, shown in Fig.~\ref{fig:smith_matching_RL}, further demonstrates that this initial design provides an excellent starting point for the final EM optimization.} \\
\indent \lastrev{The designed antenna can also be used as a means of excitation for D-band dielectric waveguides. Dielectric waveguide interconnects have attracted much attention in recent years to fill in the gap in the performance between electrical and optical links for meter-range communications~\cite{HPC_survey,HPC2,power,survey3,han,George_120G}. Such waveguides have achieved high data rates, peaking at $120~Gb/s$ in~\cite{George_120G}, where direct chip excitation of the dielectric waveguide is required. Vivaldi antennas~\cite{reynaert,reynaert2,reynaert3} and dual-patch antennas have been previously proposed to excite this waveguide~\cite{George_Neelam}. \\
\indent In this work, the proposed antenna is used to directly excite a dielectric waveguide with an M-shape refractive index profile as shown in Fig.~\ref{fig:waveguide}. The fundamental mode of this waveguide is the hybrid $HE_{11}$ mode which is a hybrid TE/TM mode. The proposed waveguide solves a fundamental multimode dispersion problem that dielectric waveguides face by mitigating the excitation of higher order modes when adiabatically tapered from the single-mode cross-section~\cite{MTT_mine}. The M-step depressed waveguide is designed and fabricated with a core of $1~mm$ width and $0.5~mm$ height (PTFE, $\varepsilon_r = 2.05$), encased by another material of 0.5 mm thickness (ePTFE, $\varepsilon_r = 1.5$). The core and cladding are wrapped with a high-density polyethylene material (HDPE) of $0.3~mm$ thickness. \\
\indent The same antenna structure proposed for the wireless case is used to excite the waveguide with the input patch impedance is plotted on Smith chart as shown in Fig.~\ref{fig:matching_waveguide}. The RL of the antenna while exciting both the wireless case and the waveguide case is shown in Fig.~\ref{fig:wireless_WG_RL}. The figure shows that the co-matching of both the wireless case and the dielectric waveguide case is possible since the initial point is simultaneously acceptable for both scenarios.}
\begin{table} [!t] 
\caption{Dimensions of the fabricated structure}
\label{table1} 
\resizebox{\linewidth}{!}{%
\centering
\begin{tabular}{c|c|c|c}
\hline
\hline
\multicolumn{2}{c|}{\textbf{Antenna Side}}& \multicolumn{2}{c}{\textbf{WR-6 Side}} \\
\hline
\hline
    \textbf{Parameter} &  \textbf{Length ($\mu m$)} & \textbf{Parameter} & \textbf{Length ($\mu m)$}\\
    \hline
     $L_{cavity}$ &  1420 & $L_{cavity2}$ &  1550\\
     \hline
      $W_{cavity}$ & 560  & $W_{cavity2}$ &  570\\
    \hline
     $L_{parasitic~patch}$ & 181  & $L_{parasitic~patch2}$ & 234 \\
     \hline
     $W_{parasitic~patch}$ & 520  &  $W_{parasitic~patch2}$ &  350 \\
     \hline
     $L_{fed~patch}$ & 400  & $L_{fed~patch2}$ & 464 \\
     \hline
     $W_{fed~patch}$ & 148  &  $W_{fed~patch2}$ &  207 \\
     \hline
     $M_{5~anti~pad}$ & 107 &  $M_{6~anti~pad}$ & 116\\
     \hline
     $M_{7~anti~pad}$ & 170 & $M_{8~anti~pad}$ & 110 \\
     \hline
      $h_{1}$ &  60 & $h_{6}$ &  40 \\
     \hline
      $h_{2}$ & 60  & $h_{7}$ &  60\\
     \hline
      $h_{3}$ &  40 & $h_{8}$ & 60\\
       \hline
      $h_{4}$ &  40 & $h_{9}$ & 40 \\
      \hline
      $h_{5}$ &  60 & $h_{10}$ & 60\\
      \hline
       &  & $h_{11}$ & 40\\
      \hline
      \hline
\end{tabular}
}
\end{table}
\begin{figure}[!t]
\centering
\includegraphics[width=\linewidth]{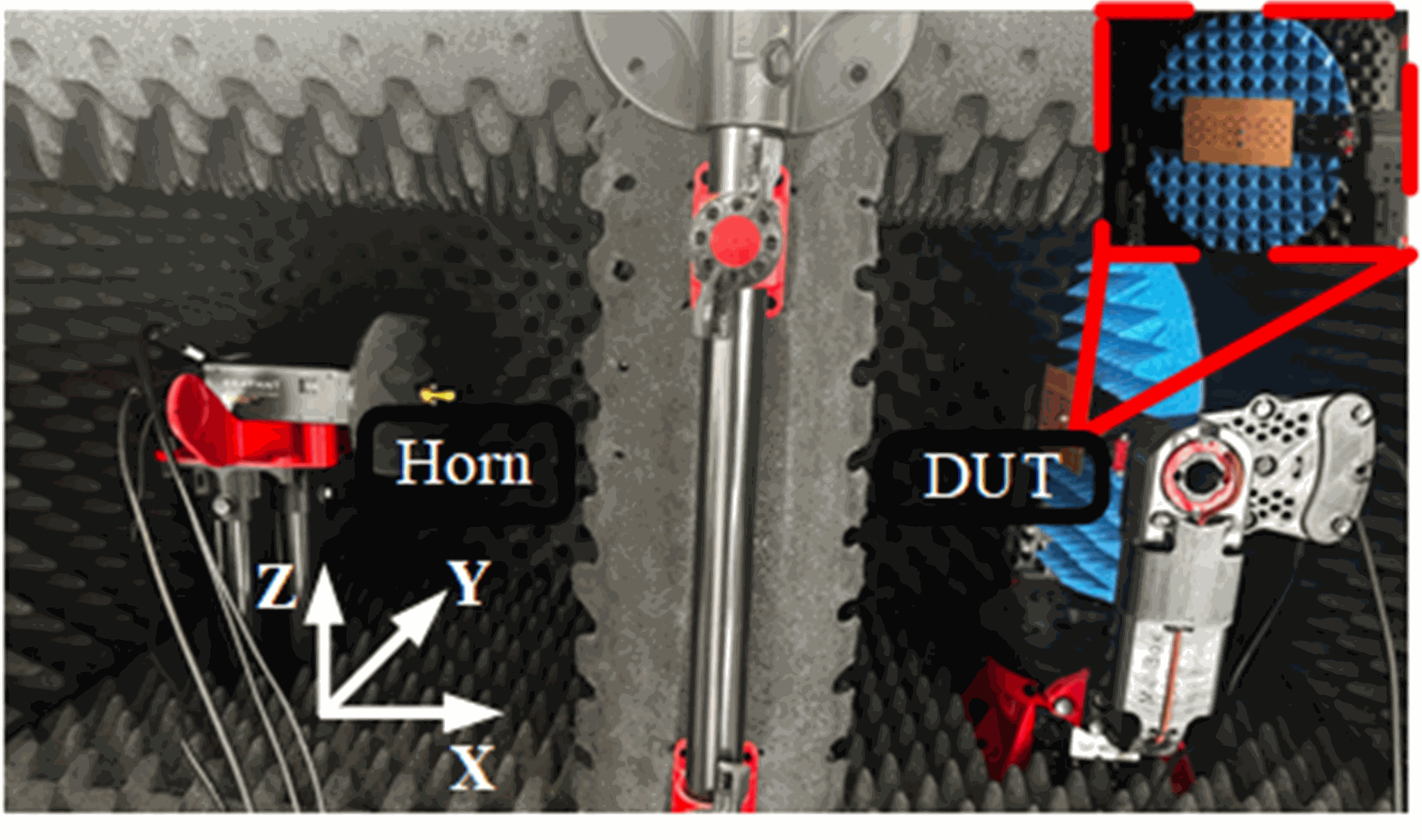}
\caption{The measurement setup of the antenna in the anechoic chamber.}
\label{fig:Setup}
\end{figure}
\begin{figure}[!t]
\centering
\includegraphics[width=\linewidth]{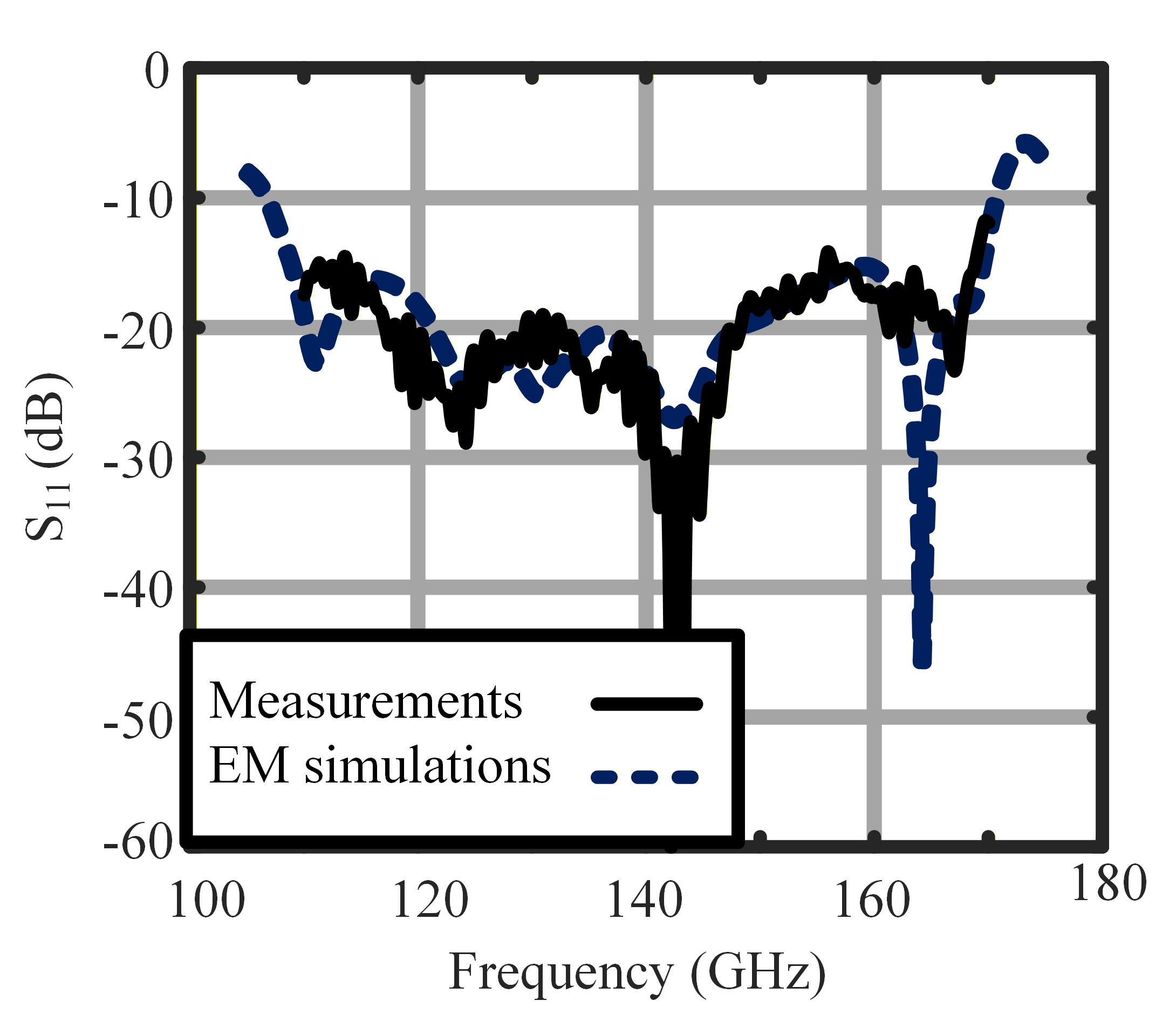}
\caption{The measured return loss of the antenna.}
\label{fig:RL}
\end{figure}
\begin{figure}[!t]
\centering
\subfloat[]{%
\centering
\includegraphics[width=0.9\linewidth]{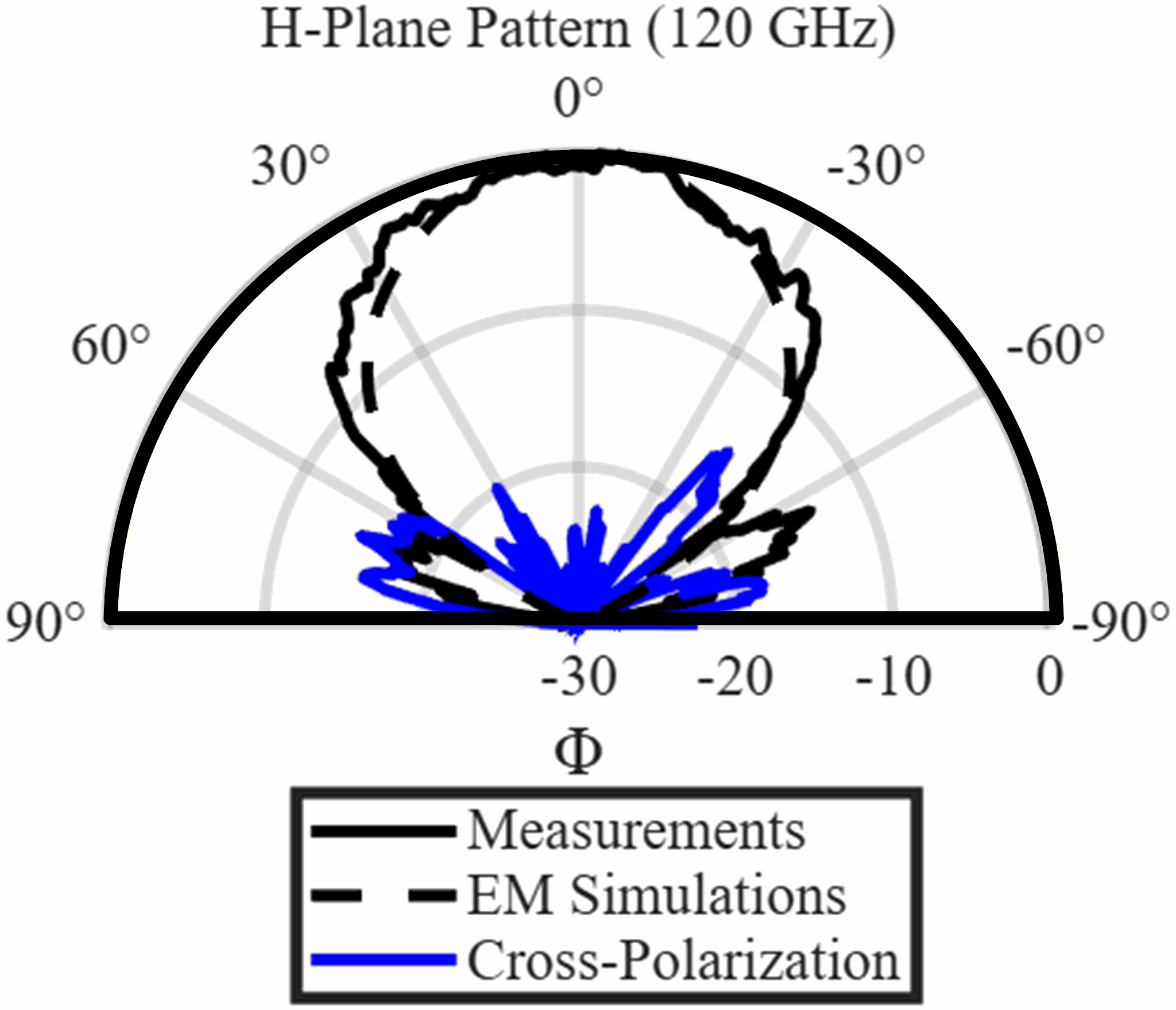}
\label{fig:H_plane_120}
}%
\\
\subfloat[]{%
\centering
\includegraphics[width=0.9\linewidth]{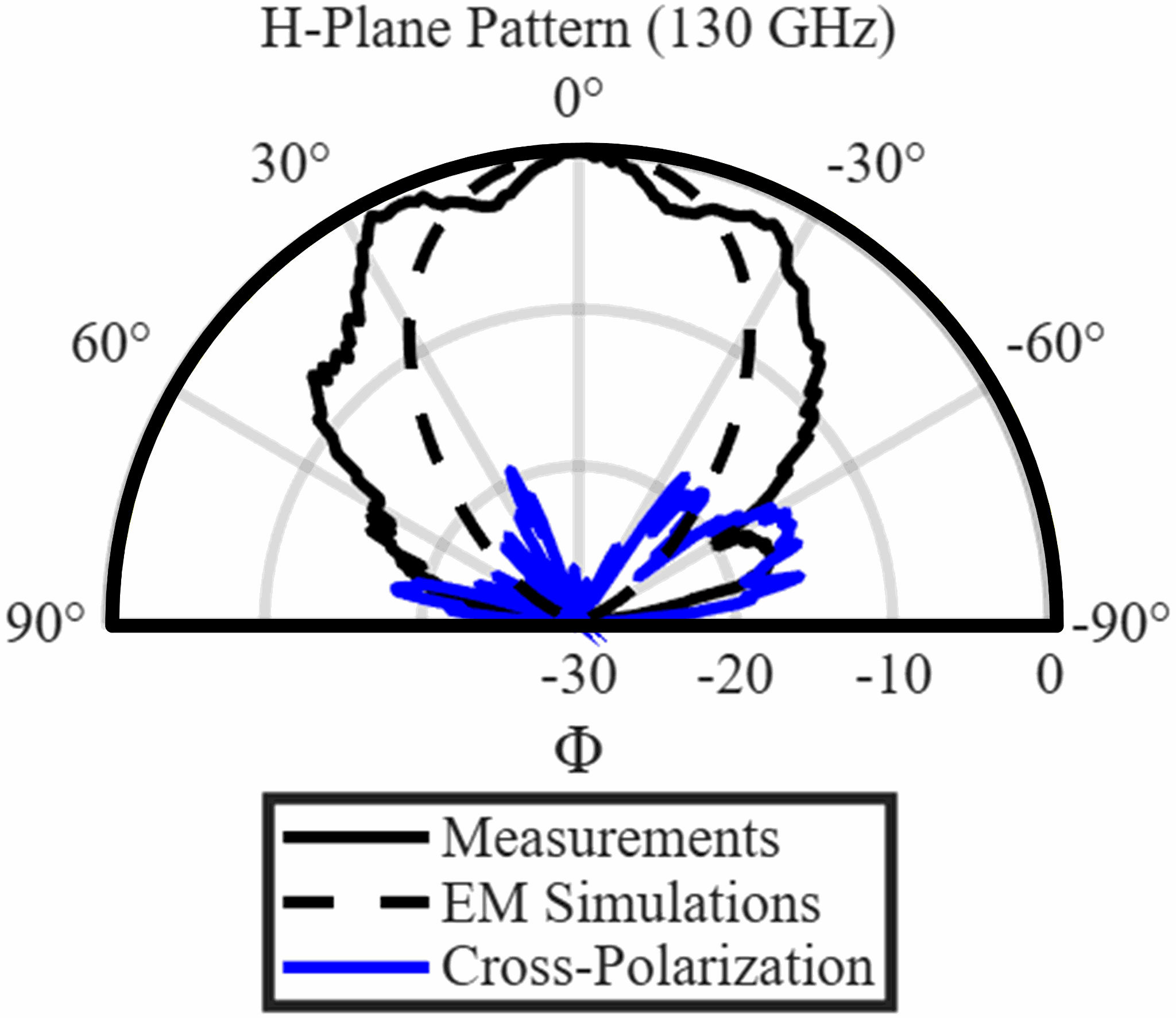}
\label{fig:H_plane_130}
}%
\\
\caption{Normalized H-plane (XY plane) pattern sweeping $\Phi$ that is defined to increase from zero on the X-axis to 90 on the Y-axis of the antenna at (a) $120~GHz$, (b) $130~GHz$. }
\label{Fig:H_plane1}
\end{figure}
\begin{figure}[!t]
\centering
\subfloat[]{%
\centering
\includegraphics[width=0.9\linewidth]{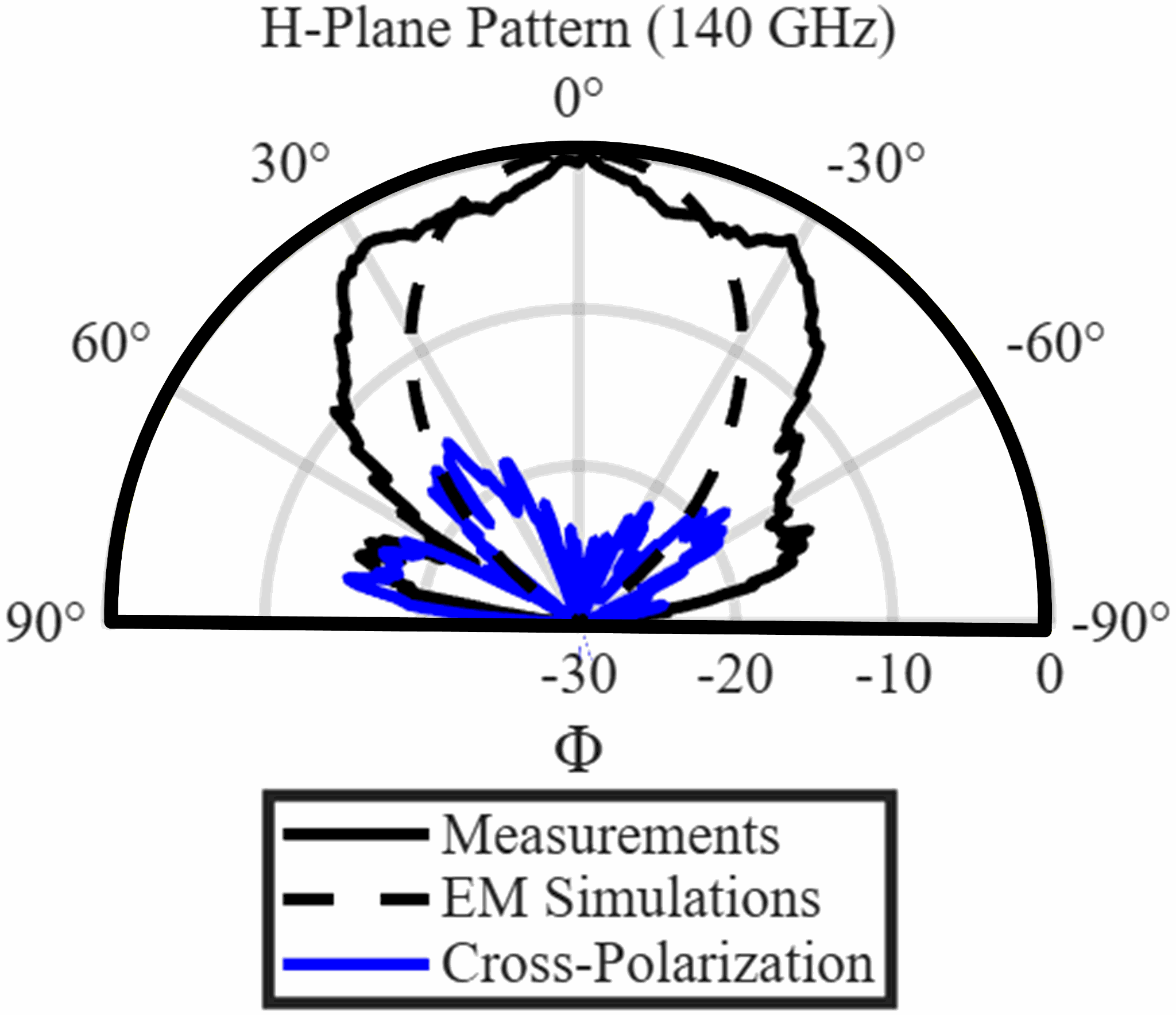}
\label{fig:H_plane_140}
}%
\\
\subfloat[]{%
\centering
\includegraphics[width=0.9\linewidth]{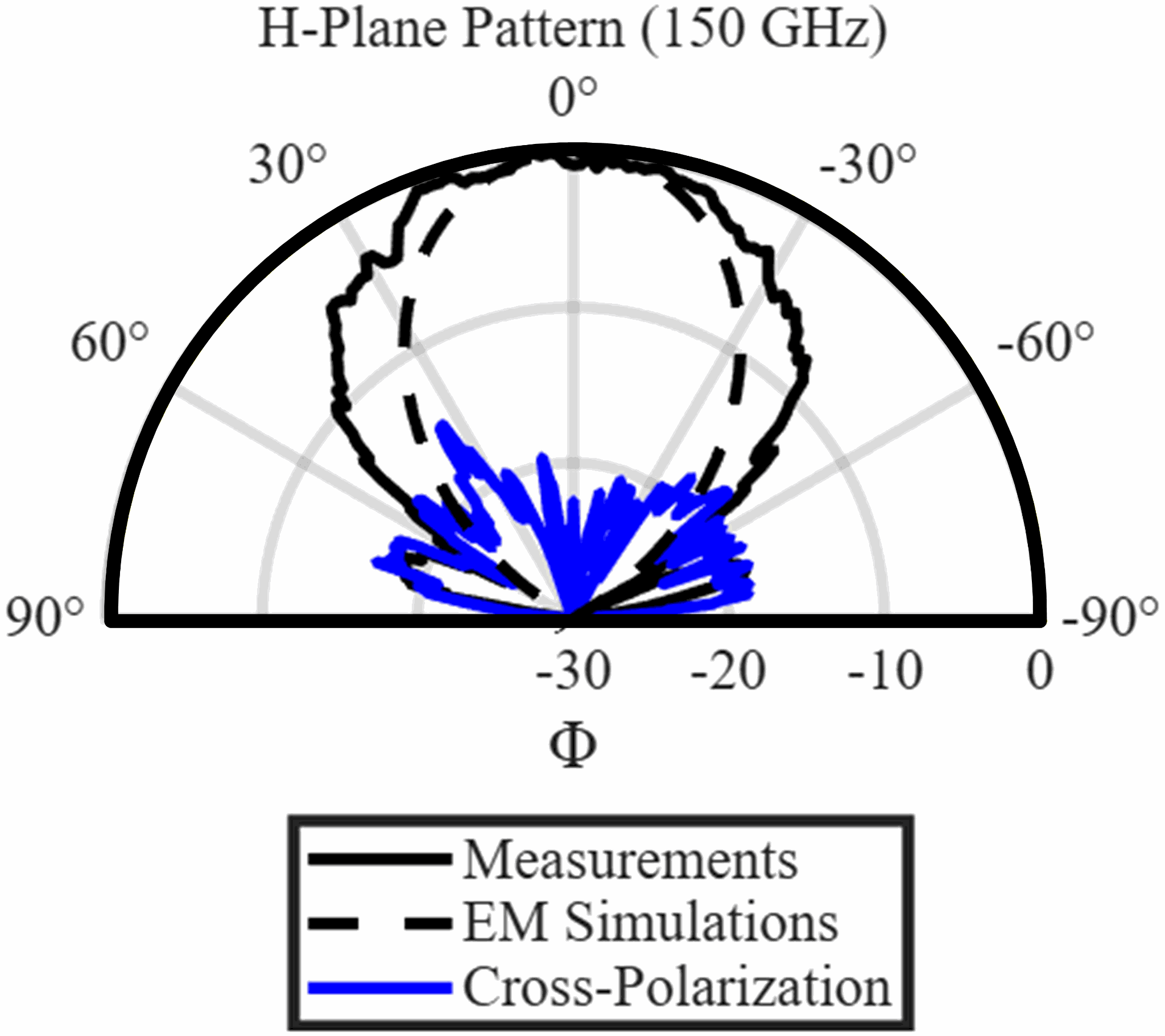}
\label{fig:H_plane_150}
}%
\\
\caption{Normalized H-plane (XY plane) pattern sweeping $\Phi$ of the antenna at (a) $140~GHz$, (b) $150~GHz$. }
\label{Fig:H_plane2}
\end{figure}
\subsection{WR-6 Transition for Measurements}
\lastrev{An auxiliary transition is built to properly interface and align the proposed antenna with the WR-6 feeding waveguide that comes out of the VNA extension module.} Similar approaches have been employed in prior work—for example, the custom-designed transition in~\cite{stacked_letter} and the commercial waveguide probe used in~\cite{payam}. \lastrev{In the proposed design, an additional stacked organic-package structure is employed to realize the transition between the radiating antenna and the WR6 waveguide. The waveguide launcher is designed following the same procedure described previously. However, an additional patch height is required to achieve the desired bandwidth. Consequently, the feed and radiating patches are positioned on the ninth and twelfth layers, respectively.
The full structure consists of two primary sections, each optimized for its respective electromagnetic interface. The upper section comprises the six metal layers extending from the topmost layer to the stripline layer. Together with the bottom ground plane, $M7$, these layers form the radiating stacked-patch antenna, which efficiently couples the guided electromagnetic energy into free space.
The lower section comprises the six metal layers associated with the WR6 interface, together with the top ground plane, $M8$. This section is designed to efficiently couple the electromagnetic energy to the fundamental mode of the WR6 waveguide while maintaining a compact and well-controlled transition geometry.
The two sections are then integrated into a single transition structure. During this process, the stripline layers $M6$ and $M9$ are removed. The feeding vias are also extended through the combined structure to ensure electrical continuity and provide a consistent electromagnetic path across the transition.
Following this integration, the two initially separate stacked structures are consolidated into a unified twelve-layer organic package. The resulting structure provides a continuous electromagnetic transition between the WR6 waveguide and the radiating stacked-patch antenna while preserving the compact multilayer nature of the proposed packaging approach. The successive integration steps and the resulting \lastrev{twelve-layer} configuration are illustrated in Fig.~\ref{Fig:structure_building}.} This configuration enables the transition and the antenna to be integrated back-to-back within a single organic package, resulting in an \lastrev{twelve-layer} structure that provides a direct interface to the WR6 waveguide of the measurement module (Fig.~\ref{fig:full_structure_cut-out}).\\
\indent Several advantages are associated with this approach as it offers low insertion loss ($<0.5~dB$), enhancing the overall system efficiency. The approach also facilitates the alignment and measurement of the antenna, as the feeding antenna interfaces directly with the waveguide using the same alignment pins and screws, eliminating the need for a challenging probe landing procedure during measurement. \\
\indent \lastrev{To characterize the transition as a standalone structure connecting the WR-6 waveguide to the radiating antenna, a back-to-back waveguide transition interfacing the WR-6 waveguide, as shown in Fig.~\ref{fig:back_to_back} is developed. This configuration integrates two identical transitions into a single package, enabling direct evaluation of the transition’s insertion loss. The corresponding measurement results, illustrated in Fig.~\ref{fig:transition}, demonstrate that the back-to-back WR-6 launcher exhibits an insertion loss below $1~dB$ (for the two-transition back to back system) at $140~GHz$, representing one of the lowest reported value for a D-band transition ($0.5~dB$ per side) to date~\cite{transition_numbers}. The design procedure for this WR-6 transition follows the same procedure discussed for the radiating antenna and the waveguide launcher. The final dimensions of the cavity, patches, via pads, and antipads of the full structure are given in Table~\ref{table1}}.
\begin{figure}[!t]
\centering
\subfloat[]{%
\centering
\includegraphics[width=0.9\linewidth]{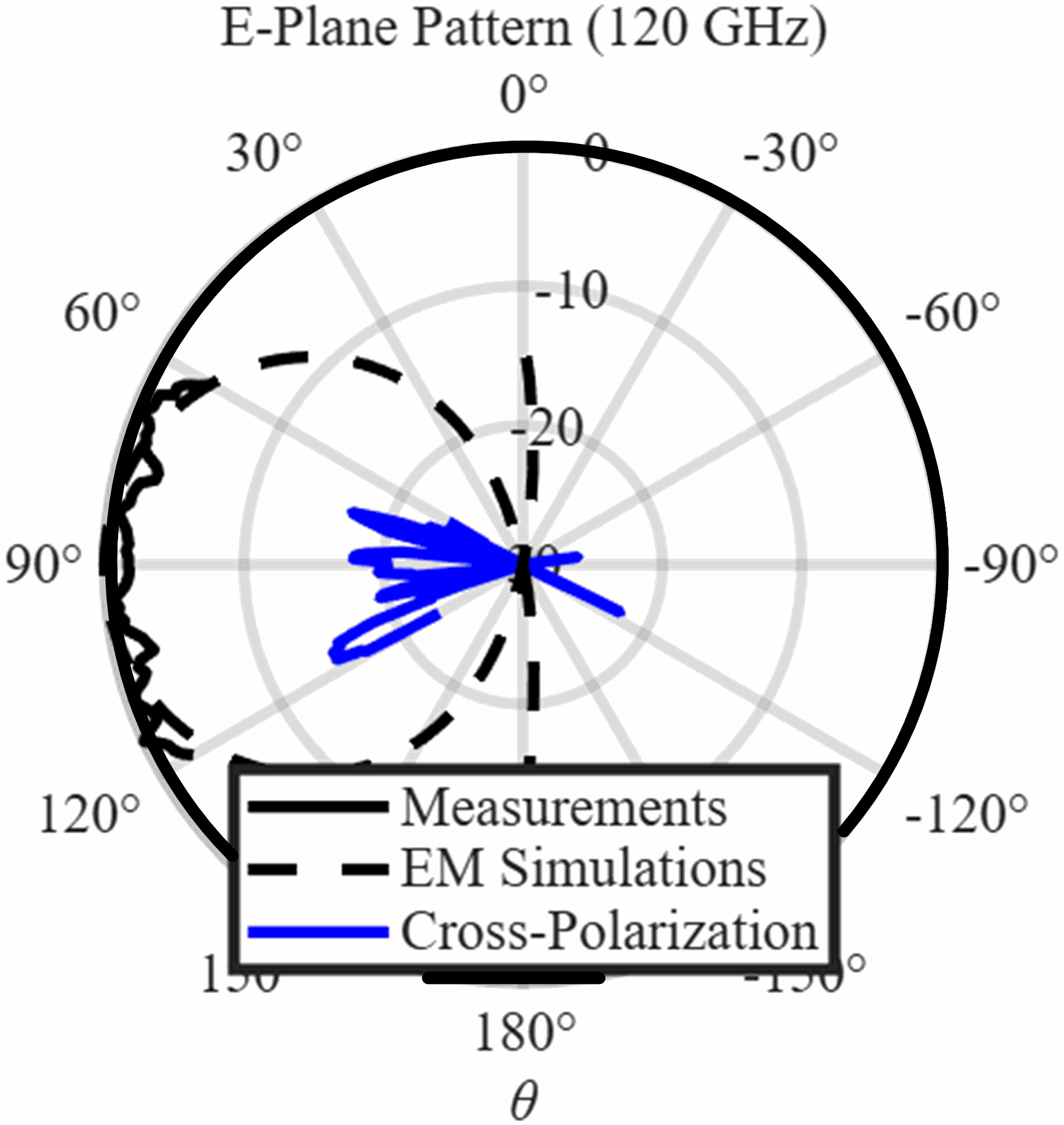}
\label{fig:E_plane_120}
}%
\\
\subfloat[]{%
\centering
\includegraphics[width=0.9\linewidth]{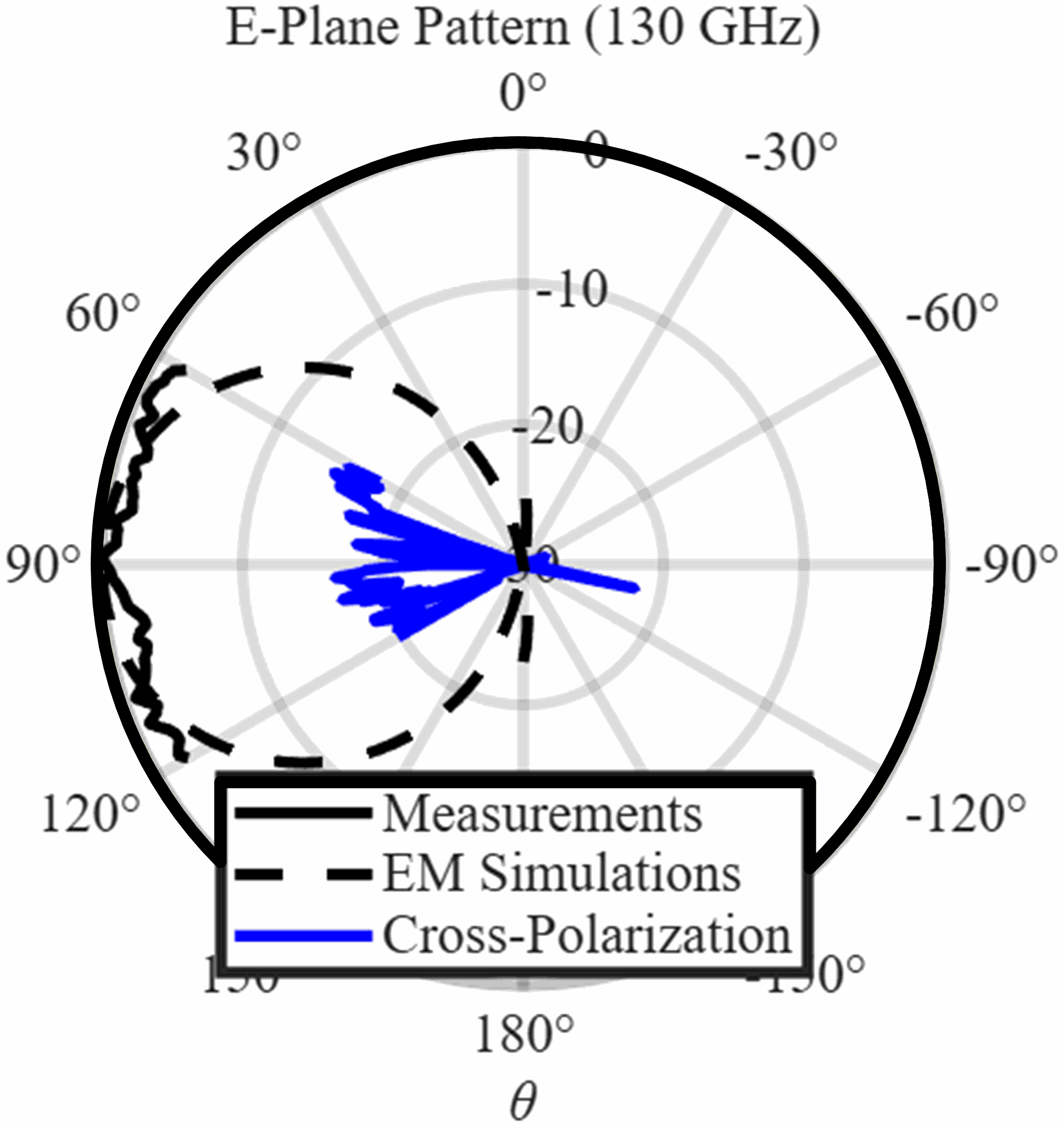}
\label{fig:E_plane_130}
}%
\caption{Normalized E-plane (XZ plane) pattern sweeping $\theta$ of the antenna at (a) $120~GHz$, (b) $130~GHz$.}
\label{Fig:E_plane1}
\end{figure}
\begin{figure}[!t]
\centering
\subfloat[]{%
\centering
\includegraphics[width=0.9\linewidth]{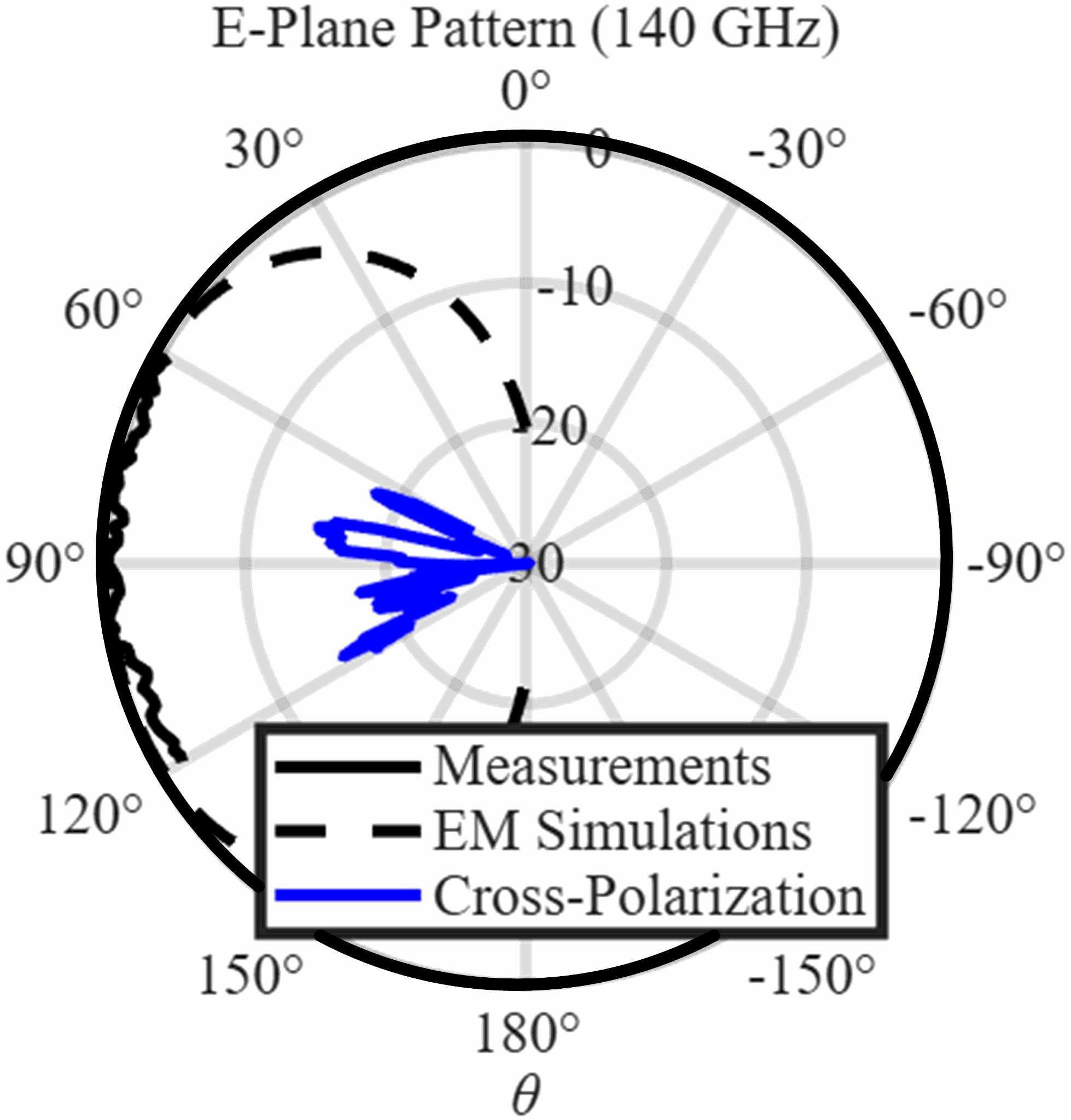}
\label{fig:E_plane_140}
}%
\\
\subfloat[]{%
\centering
\includegraphics[width=0.9\linewidth]{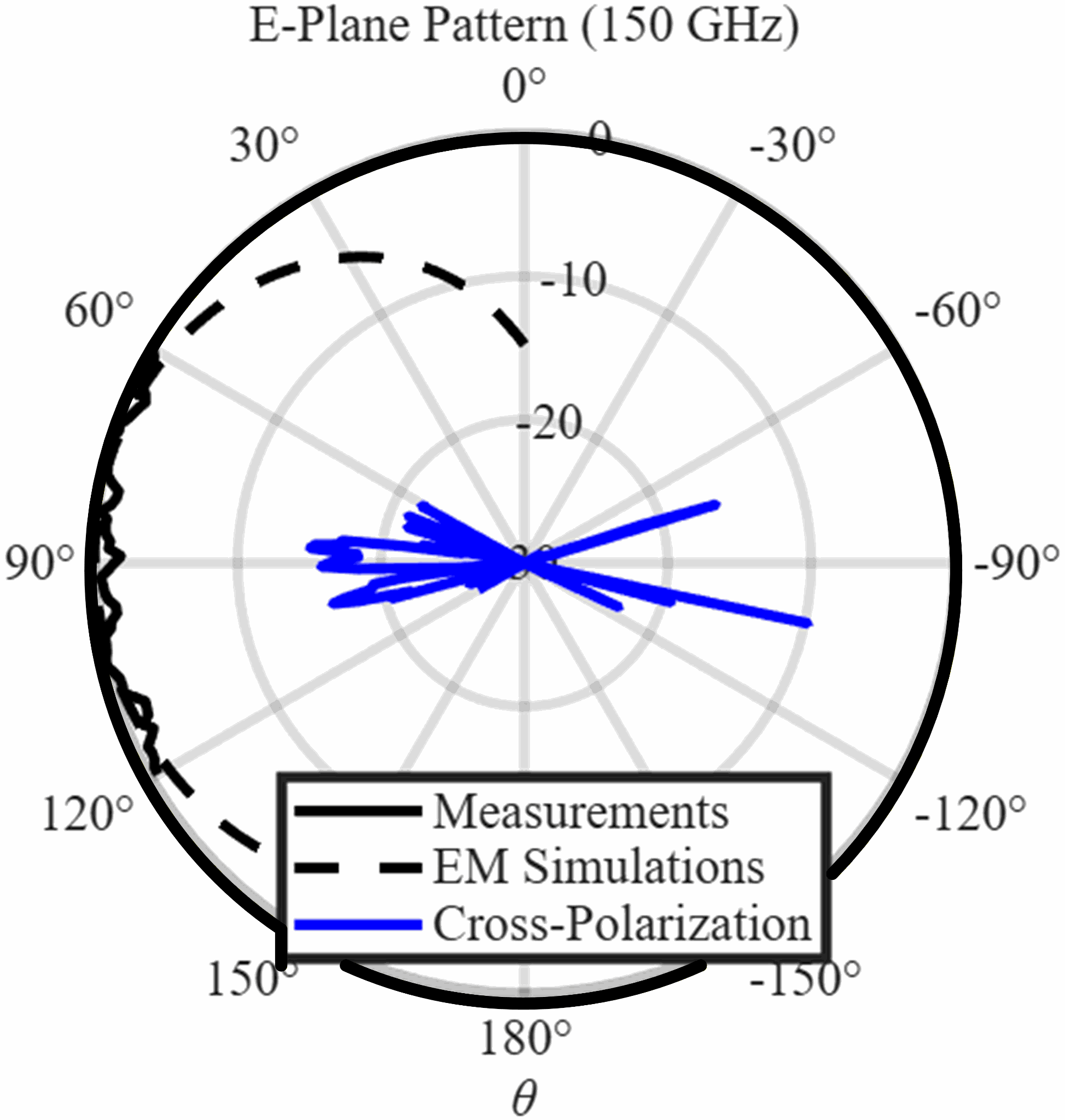}
\label{fig:E_plane_150}
}%
\caption{Normalized E-plane (XZ plane) $\theta$ pattern of the antenna at (a) $140~GHz$, (b) $150~GHz$.}
\label{Fig:E_plane2}
\end{figure}
\begin{figure}[!t]
\centering
\includegraphics[width=\linewidth]{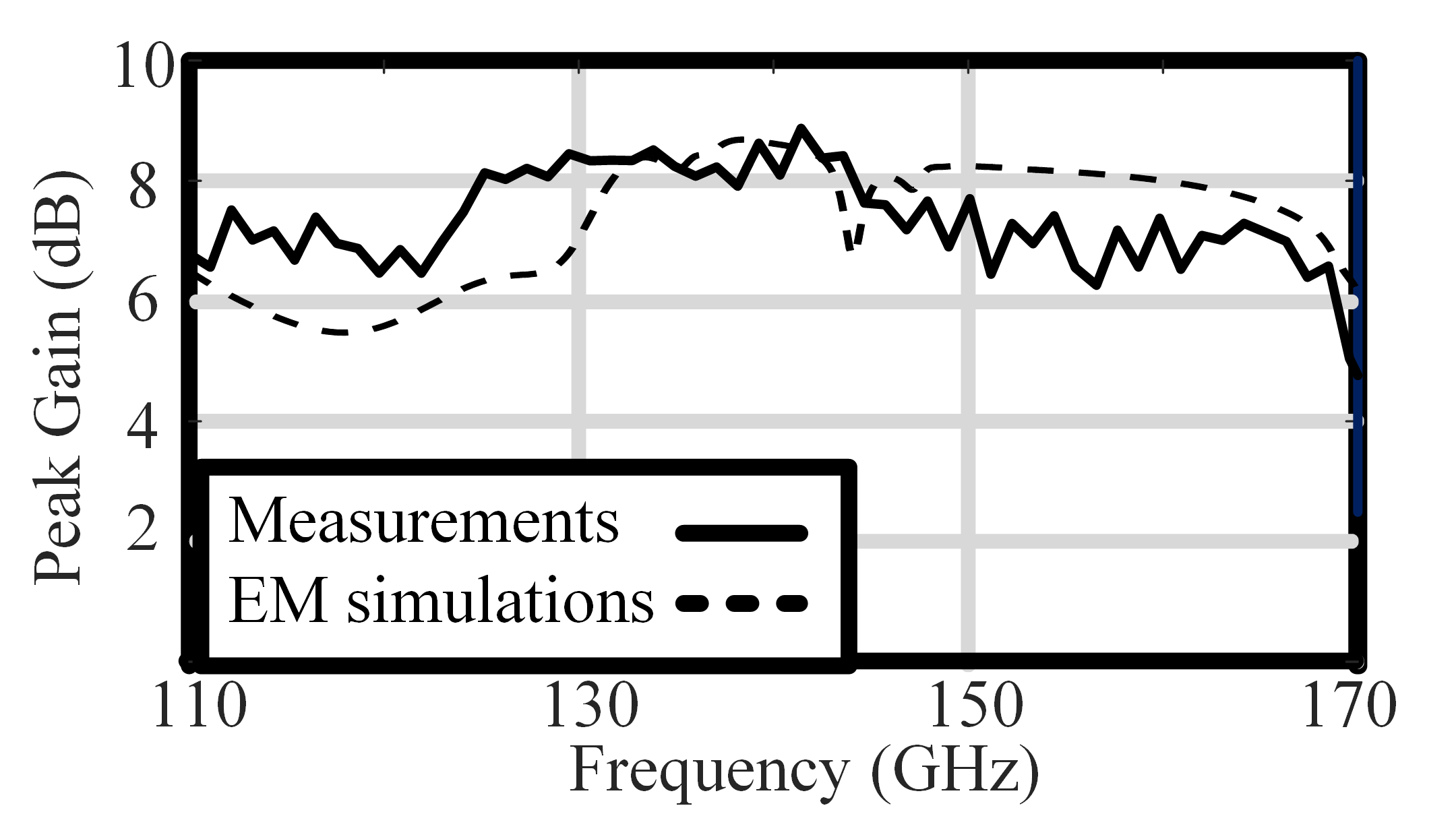}
\caption{The measured peak gain ($\theta = 90$ and $\Phi = 0$) of the antenna.}
\label{fig:Gain}
\end{figure}
\begin{figure}[!t]
\centering
\includegraphics[width=\linewidth]{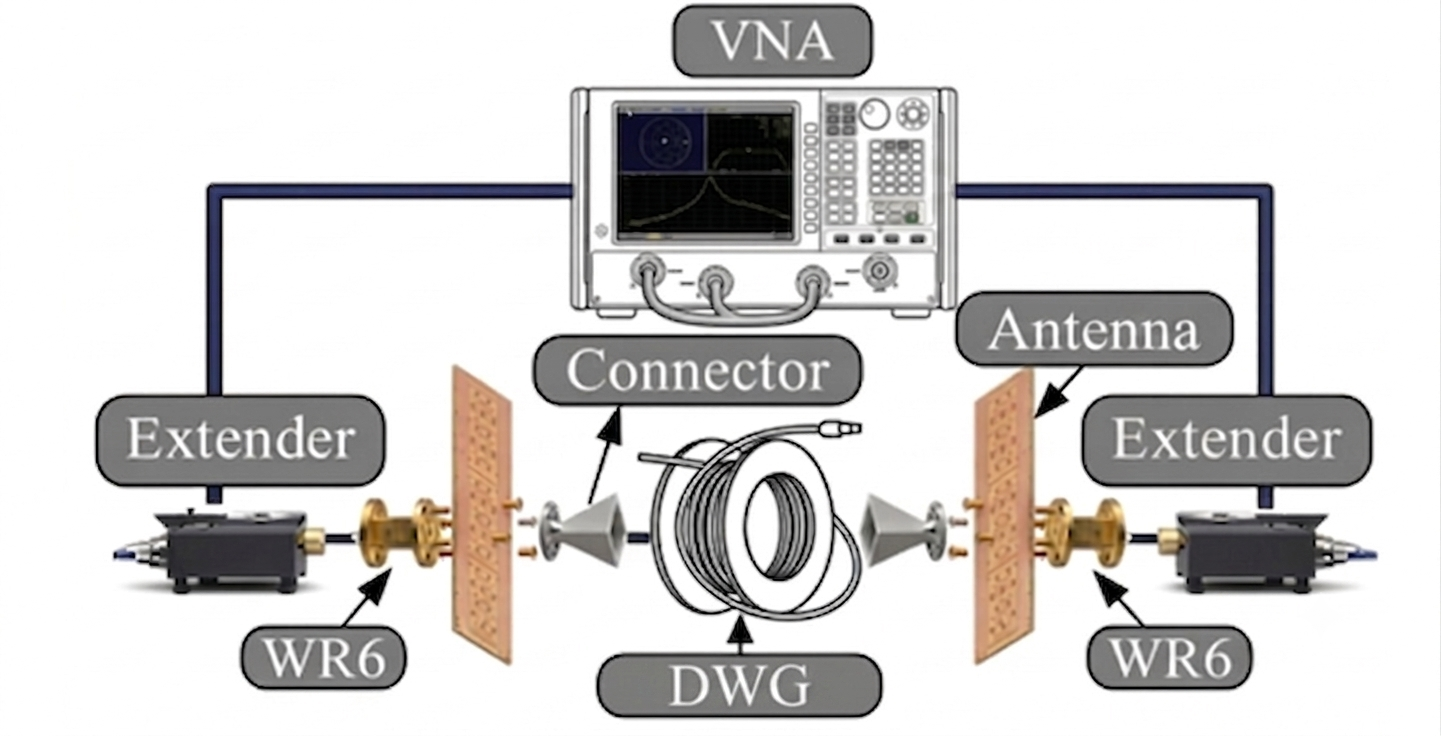}
\caption{The measurement setup of the WR6 stacked-patch launcher feeding the proposed antenna, and feeds a 1-m dielectric waveguide.}
\label{fig:waveguide_structure}
\end{figure}
\begin{figure}[!t]
\centering
\includegraphics[width=\linewidth]{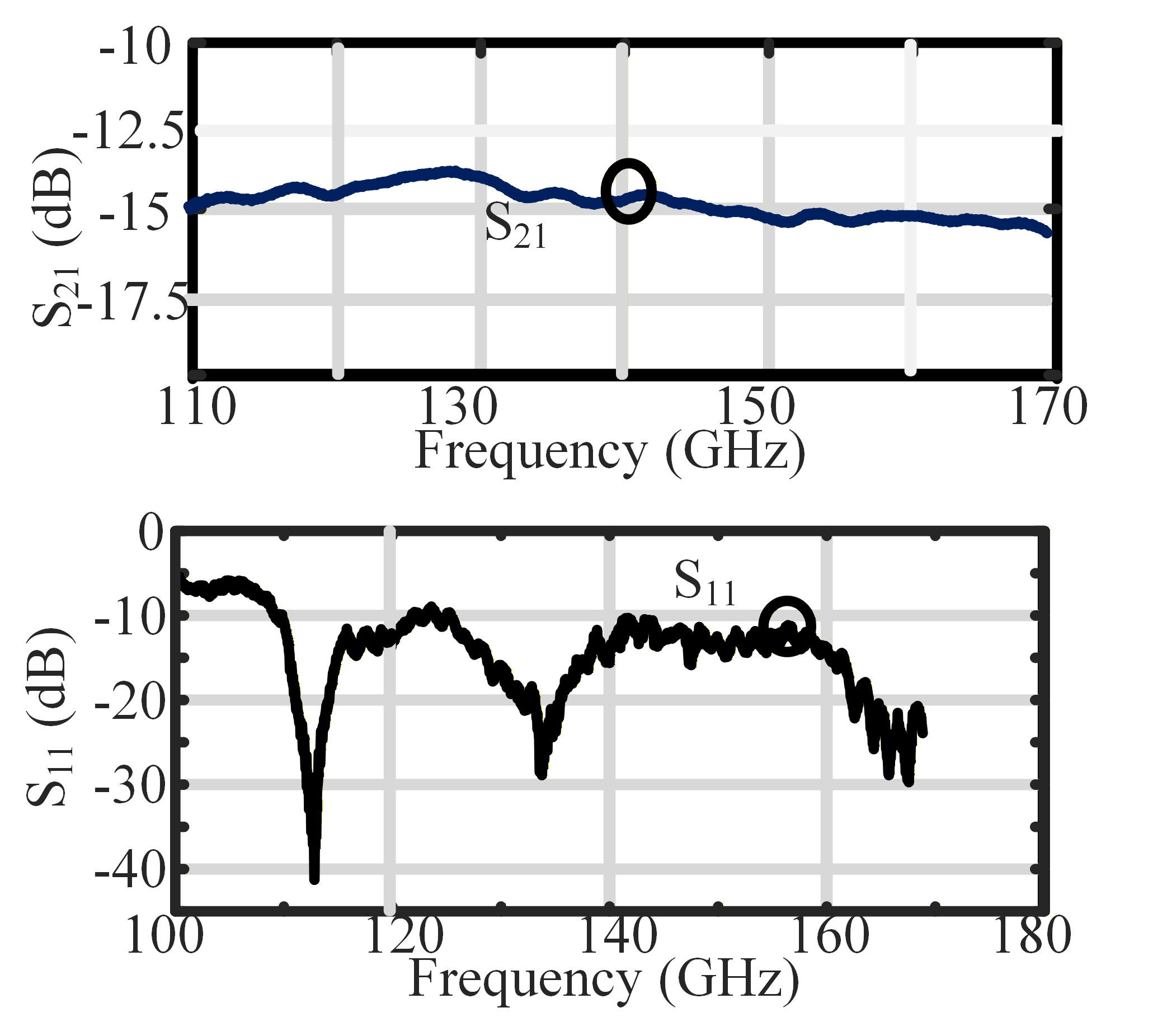}
\caption{The measurement response of 1-m dielectric waveguide fed by two back-to-back stacked-patch launcher transitions. Note that one transition per waveguide side includes two launchers, therefore there are four stacked-patch launchers in the measured response.}
\label{fig:waveguide_feed}
\end{figure}
\section{Measurement Results}
\lastrev{The proposed design, illustrated in Fig.~\ref{fig:full_structure_cut-out} is fabricated on an twelve-layer organic packaging technology as shown in Fig.~\ref{fig:fab_antenna}. The design incorporates alignment holes that precisely interface with the WR6 flange alignment pins, ensuring accurate mechanical alignment between the package and the VNA-extender interface.}
\begin{table} [!t] 
\caption{Comparison against prior work}
\label{table2} 
\begin{tabularx}{\linewidth}{>{\centering\arraybackslash}X|>{\centering\arraybackslash}X|>{\centering\arraybackslash}X|>{\centering\arraybackslash}X|>{\centering\arraybackslash}X|>{\centering\arraybackslash \columncolor{lightgray}}X}
\hline
\hline
\centering
& AWPL`19\newline~\cite{dipole} & TAP `22\newline~\cite{payam} & AWPL`24\newline~\cite{stacked_letter} &  TCP`16\newline~\cite{stacked_third} & \textbf{This Work} \\
\hline
Antenna Type & Patch &\multicolumn{4}{c}{\textbf{Dual stacked patch}}\\
\hline
\lastrev{Operating Freq. (GHz)} & 150-180 & 90-128.5 &\multicolumn{3}{c}{\textbf{110-170}} \\
\hline
$10-dB$ RL\newline FBW\newline(GHz) & 10  & 38.5 & 40 & 15 & \textbf{$>$ 60} \\
\hline
Gain\newline(dBi)   &  5 & 7.95 & 7.5$^1$ & N/A$^2$ & \textbf{8.5}\\
\hline
Gain Variation\newline(dBi)   & 3 & 3.8 & 3 & 6 & \textbf{3.5}\\
\hline
Cross polarization\newline Coupling &\multicolumn{2}{c|}{N/A} & -10 & -15 & \textbf{-20}\\
\hline
$3-dB$\newline Beamwidth  & N/A & 10 & 20 & 15 &\textbf{35}\\
\hline
Transition\newline Loss &\multicolumn{2}{c|}{N/A} & 7 & N/A &\textbf{1}\\
\hline
Size\newline(mm$^2$)   & 1 & 17.6$^3$ &  2.56$^4$ & N/A &\textbf{0.86}\\
\hline
Metal layers   & 7 & 4 & 6 & 3 & \textbf{6}\\
\hline
\lastrev{Waveguide Excitation}   & \multicolumn{4}{c|}{N/A}& \textbf{yes}\\
\hline
\hline
\end{tabularx}
\\
   \footnotemark[1]{Single element gain.} \hspace{15mm}
   \footnotemark[2]{Gain of the individual element is not reported.} \newline
   \footnotemark[3]{Calculated based on the smallest wavelength.}\newline
    \footnotemark[4]{Calculated from the array dimensions.}
\end{table}
\lastrev{Fig.~\ref{fig:fab_antenna} shows the antenna structure with the coordinate system used for the measurements. According to this coordinate system, the antenna is predicted to radiate along the X-axis with the H-plane and E-plane cuts being the X-Y and X-Z planes respectively.} The antenna gain and radiation pattern are measured inside an anechoic chamber with electronic beam alignment as shown in Fig.~\ref{fig:Setup}. \lastrev{The measured RL response is shown in Fig.~\ref{fig:RL}. The RL remains below $15~dB$ across the entire D-band due to the four in-band resonances as explained in Section II. This measurement demonstrates the widest reported stacked-patch antenna bandwidth in the literature to the best of our knowledge.} \\
\indent The H-plane (X-Y plane) pattern measurements, are plotted as the boresight angle ($\phi$), measured as the angle from the X-axis, varies from $-90$ to $90^\circ$. The measurements are plotted at 120 and $130~GHz$ as shown in Fig.~\ref{Fig:H_plane1}. The $3-dB$ beamwidth at $120~GHz$ is $42^\circ$, while it is $36^\circ$ at $130~GHz$. The plots also show more than $20~dB$ and $30~dB$ of co- to cross-polarization rejection at the maximum radiation angle for the $120$ and $130~GHz$ plots, respectively. The pattern is also drawn for $140$ and $150~GHz$ respectively in Fig.~\ref{Fig:E_plane2} showing a $3-dB$ beamwidth of $35^\circ$ each. The peak co- to cross-modal isolation of the antenna is more than $35~dB$ in both cases. \lastrev{The measurements show a good agreement with the EM-simulations, especially at $120~GHz$. Discrepancies at higher frequencies are attributed to mechanical non-idealities and fabrication tolerances.}
\\ \indent For the E-plane measurements, X-Z plane, illustrated in Fig.~\ref{fig:fab_antenna}, the measurements are done by sweeping the boresight angle $\theta$ from $60$ to $120$, i.e., $\pm 30^\circ$ around the maximum $\theta = 90$. Measurements are conducted for 4 different frequencies ($120$, $130$, $140$, and $150~GHz$) as illustrated in Fig.~\ref{Fig:E_plane1} and~\ref{Fig:E_plane2}. \lastrev{The measurements demonstrate a co- to cross-polarization isolation of more than $20~dB$ across the entire sweep and agree very well with the EM simulations.}
\\ \indent \lastrev{The peak gain of the antenna ($\theta = 90$ and $\phi = 0$) is measured across the entire D-band for both the co- and cross-polarization configurations. The measurements are done using a standard horn antenna with a reference gain of $20~dBi$ \lastrev{and a gain variation of $2~dB$} placed at $50~cm$ distance from the designed antenna inside an anechoic chamber, as illustrated in Fig.~\ref{fig:Setup}}. The maximum gain is $8.5~dBi$ with a gain variation of $3.5~dBi$ in the entire D-band. \lastrev{This gain variation, although measured across a much wider bandwidth ($60~GHz$), still compares favorably as an absolute number to other state-of-the-art D-band antennas with smaller bandwidths.} \\
\indent Table~\ref{table2} compares the proposed antenna with other state-of-the-art D-band antennas. The antenna exhibits the widest measured bandwidth reported as indicated by the beamwidth gain product, along with very low losses associated with the designed WR6 transition. The simulated efficiency is higher than 90\% with a sub-10\% variation across the entire operating bandwidth. This efficiency is verified by the strong agreement between gain measurements and simulations. The antenna also has the smallest size (area-wise) and exhibits the best co-to-cross-polarization isolation. 
\\ \indent \lastrev{The designed D-band antenna is also used to feed the waveguide proposed in~\cite{MTT_mine} as depicted in Fig.~\ref{fig:waveguide_structure}.} \lastrev{The waveguide is mounted on the antenna using a plastic pyramidal shape connector made of a plastic material. This pyramidal connector has a base that can interface directly into the alignment pins of the WR-6 flange. This ensures perfect alignment of the WR-6-antenna-connector structure. The waveguide itself is inserted in the connector and held in place via a micro-machined low-density (foam) polyethylene filler that adds minimal loss to the overall system.} To our knowledge, this is the first antenna to demonstrate high measured peformance for both wireless and dielectric waveguide interconnect cases. \lastrev{The antenna excites the fundamental mode $HE_{11}$ in a single-mode waveguide. The input single-mode waveguide is tapered into a $1~m$ long multi-mode waveguide.} The $10~dB$ RL bandwidth of the waveguide link is $60~GHz$ from $110$--$170~GHz$. The system exhibits a return loss of $14.7~dB$ as illustrated in Fig.~\ref{fig:waveguide_feed} for the whole system, in which the waveguide channel itself contributes $7~dB/m$ loss, and the transition has a simulated insertion loss of roughly $3.5~dB$ per transition on each end of the waveguide. Note that one transition per waveguide side includes two launchers, therefore there are four stacked-patch launchers in the measured response of Fig.~\ref{fig:waveguide_feed}.
\section{Conclusion}
The paper presented a dual-stacked patch antenna operating in the D-band. The antenna is implemented on a 7-layer organic packaging technology, exhibiting the widest reported RL bandwidth, covering the full $60~GHz$ of the D-band with better than $-15~dB$ RL bandwidth. A WR6 waveguide-to-antenna transition is designed and integrated, exhibiting a very small insertion loss ($<~1$ dB). The full structure is integrated together, back-to-back, and measured inside an anechoic chamber, achieving a measured gain of $8.5~dBi$ with a gain variation of $3.5~dBi$ across the bandwidth. The same design is also used as a transition to a dielectric waveguide without sacrificing the RL bandwidth and with an IL of $3.5~dB$. The antenna measurements match closely with the predicted simulation results. 
\section*{Acknowledgment}
The authors gratefully acknowledge Eravant Inc. for conducting the antenna parameter measurements in its anechoic chamber.

\begin{IEEEbiography}[{\includegraphics[width=1in,height=1.25in,clip,keepaspectratio]{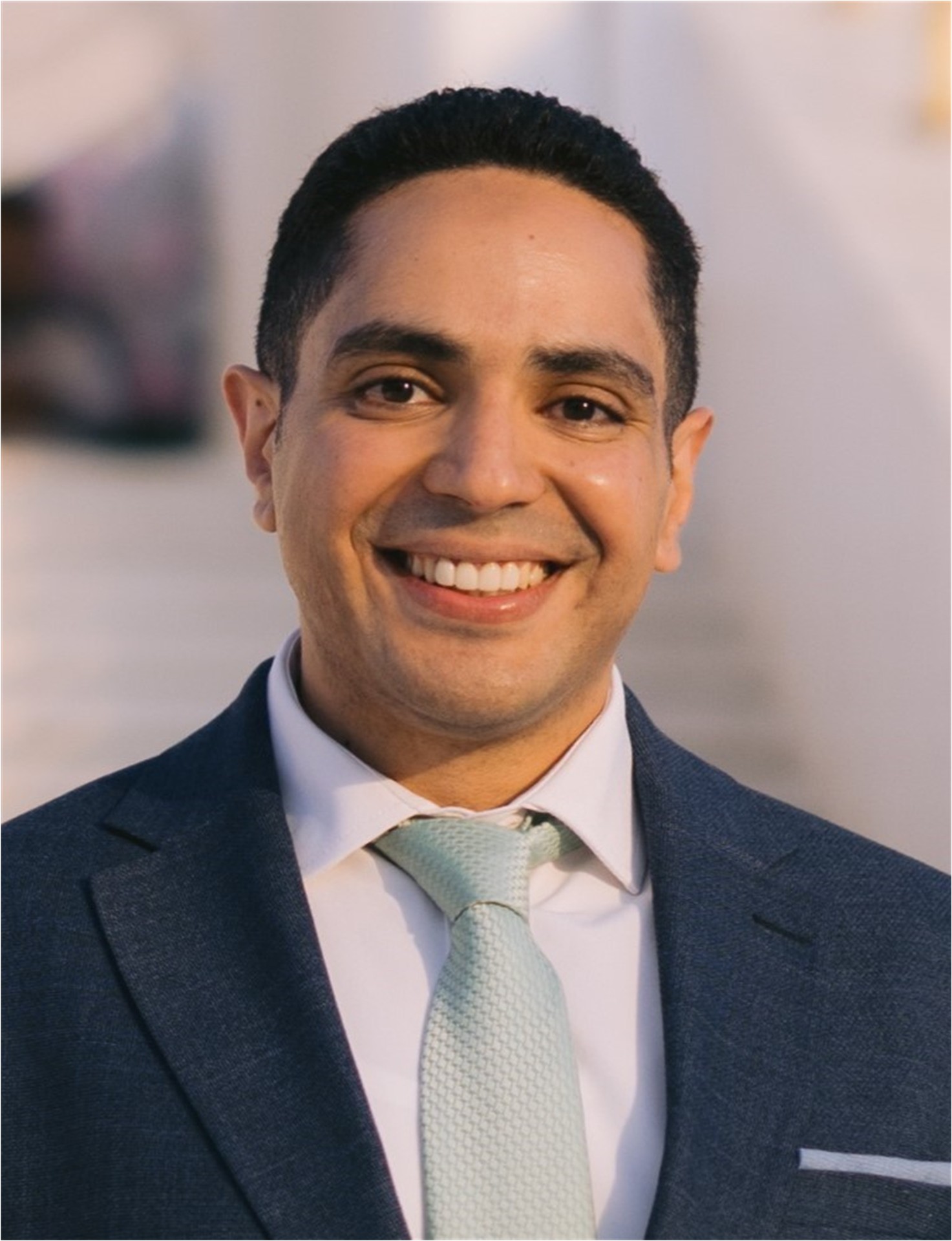}}]{Mohamed Elsawaf}{\space}(Graduate Student Member, IEEE) received the B.Sc. and M.Sc. degrees with honors in Electrical Engineering from Ain Shams University, Cairo, Egypt, in 2018 and 2021, respectively. Following his graduation, he joined Siemens EDA (formerly Mentor Graphics), where he worked for three years as a Software Development Engineer in the Device Modeling Team. Concurrently, he served as a Research Assistant at the Microwave and Antenna Research Lab (MARL), Ain Shams University.
Dr. Elsawaf earned his Ph.D. degree in Electrical and Computer Engineering from the University of Southern California (USC), Los Angeles, CA, USA, in May 2026. He is currently a Senior Staff Design Engineer at Marvell Technology, specializing in high-speed interconnect design.
His research interests span passive and active microwave and terahertz (THz) components, with applications in advanced communication systems and biomedical technologies. Dr. Elsawaf was awarded the USC Graduate Research Fellowship in 2021.  
\end{IEEEbiography}
\begin{IEEEbiography}[{\includegraphics[width=1in,height=1.25in,clip,keepaspectratio]{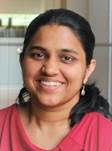}}]{Neelam Prabhu Gaunkar}{\space}(Member, IEEE)  received her Master’s (2014) and PhD (2018) in electrical engineering from Iowa State University (ISU). Since 2018 she has worked at Intel Foundry and Manufacturing, Technology Research in Chandler, AZ, as a semiconductor research engineer. Her research interests include applied electromagnetics, millimeter wave packaging, biomagnetism, signal integrity, and high-speed systems. She has published over 30 peer-reviewed articles and co-authored over 40 patent applications.   
\end{IEEEbiography}
\begin{IEEEbiography}[{\includegraphics[width=1in,height=1.25in, clip,keepaspectratio]{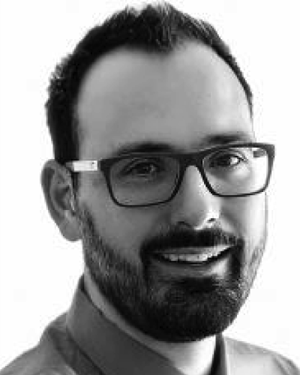}}]{Georgios C. Dogiamis}{\space}(Member, IEEE) received the Dipl.-Ing. degree in electrical and computer engineering from the National Technical University of Athens, Athens, Greece, in 2006, the M.Sc. degree in electrical engineering from California Institute of Technology, Pasadena, CA, USA, in 2009, and the Ph.D. degree in electrical engineering from the University of Duisburg-Essen, Duisburg, Germany, in 2014.\\
He was with the Informatics Department, Hellenic Army General Staff, Athens, from 2007 to 2009. Since 2009, he has been a Staff Researcher with the University of Duisburg-Essen and the Fraunhofer Institute (IMS), Duisburg. He joined the Technology Research team, Intel Corporation, Chandler, AZ, USA, in 2014, where he became a Principal Engineer and a Director of the Systems \& Advanced Packaging Process Engineering team. Since June 2025 he is the Sr. Director of Technology Research at Deca Technologies, Tempe, AZ, USA. His current research interests include advanced semiconductor packaging and assembly techniques for high performance computing systems, high bandwidth millimeter-wave systems, and co-packaged photonic systems.
\end{IEEEbiography}
\begin{IEEEbiography}[{\includegraphics[width=1in,height=1.25in,clip,keepaspectratio]{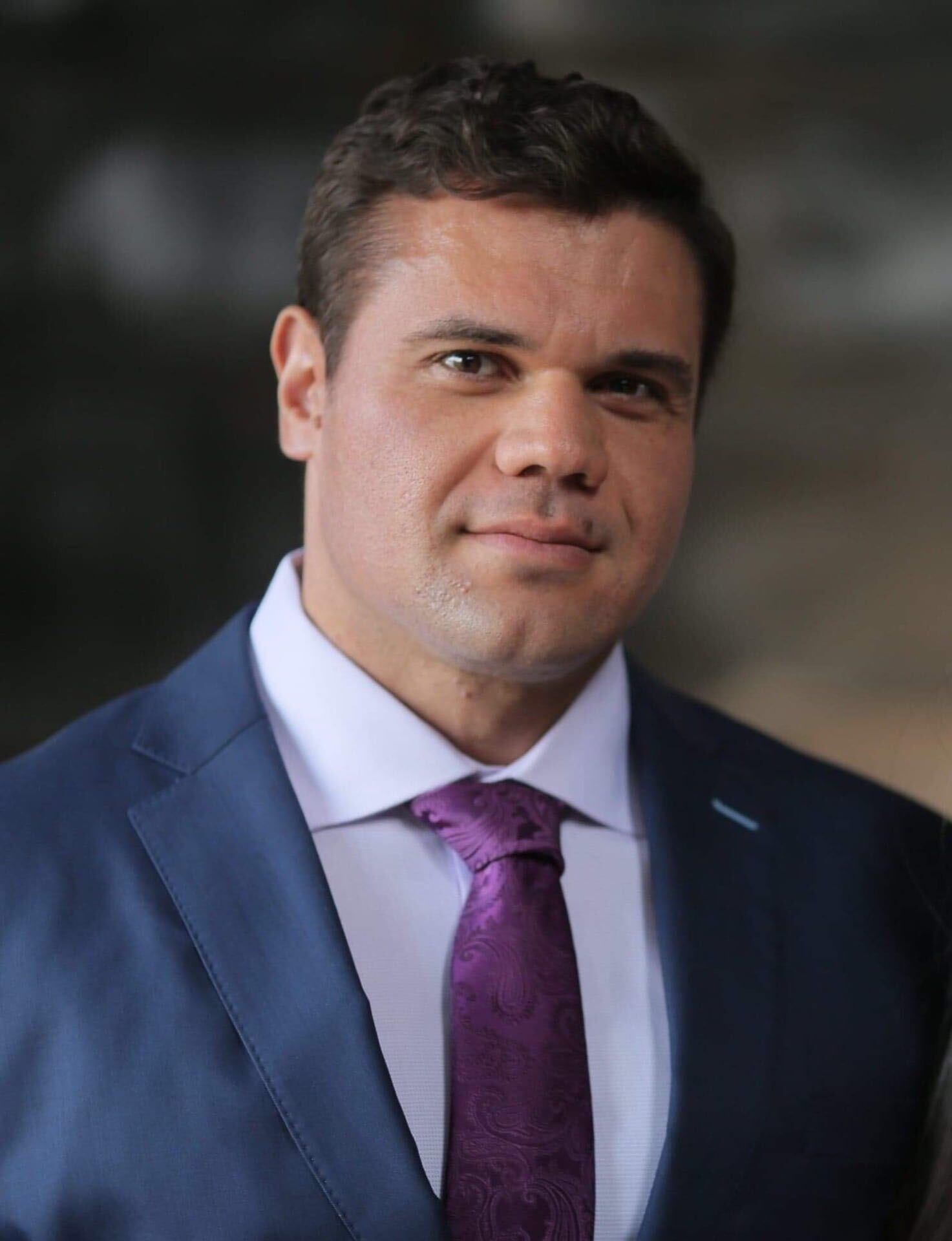}}]{Constantine Sideris}{\space}(Senior Member, IEEE) received the B.S., M.S., and Ph.D. degrees (Hons.) from the California Institute of Technology (Caltech), Pasadena, CA, USA, in 2010, 2011, and 2017, respectively. He was a Visiting Scholar with the Berkeley Wireless Research Center, Berkeley, CA, from 2013 to 2014. He was a postdoctoral scholar with the Department of Computing and Mathematical Sciences, Caltech, from 2017 to 2018, working on integral equation methods for electromagnetics. He is currently an Associate Professor of Electrical Engineering at Stanford University, Stanford, CA. Previously, he was an Assistant Professor at the University of Southern California (USC) from 2018 to 2025 and an Associate Professor at USC from 2025 to 2026. His research interests include RF and millimeter-wave integrated circuits for bioelectronics and wireless communications, applied electromagnetics, and computational electromagnetics for antenna design and nanophotonics. Dr. Sideris was a recipient of the ONR YIP Award in 2023, the NSF CAREER Award in 2021, the AFOSR YIP Award in 2020, the Caltech Leadership Award in 2017, and the NSF Graduate Research Fellowship in 2010. 
\end{IEEEbiography}
\end{document}

% Insert where needed to balance the two columns on the last page with
% biographies
%\newpage

% You can push biographies down or up by placing
% a \vfill before or after them. The appropriate
% use of \vfill depends on what kind of text is
% on the last page and whether or not the columns
% are being equalized.

%\vfill

% Can be used to pull up biographies so that the bottom of the last one
% is flush with the other column.
%\enlargethispage{-5in}

% that's all folks
\end{document}